\documentclass{appolb}
\usepackage{graphicx}
\usepackage{epsfig}

\begin{document}
\title{High Energy Evolution - The Wave Function Point of View%
\thanks{Presented at XLV Cracow School of Theoretical Physics, Zakopane June 2005}%
}
\author{Alex Kovner
\address{Physics Department,
University of Connecticut,\\ 2152 Hillside Road, Storrs, CT 06269, USA}
}
\maketitle
\begin{abstract}
These lectures discuss aspects of high energy evolution in QCD. This includes the derivation of the JIMWLK equation, basic physics of its solutions and recent work on inclusion of Pomeron loops. The entire discussion is given in the Hamiltonian framework which gives direct access to the evolution of hadronic wave function under Lorentz boost.
\end{abstract}


\section {Introduction}
These lectures deal with the description of hadronic scattering at high energies, and more specifically with the
evolution of the scattering matrix with energy\cite{pomeron},\cite{mv},\cite{reggeon}. It has been a very lively and active field for the last 10 years\cite{balitsky},\cite{JIMWLK},\cite{cgc},\cite{kovchegov} and especially lately there has been a 
new burst of activity. The latest developments are related to what is sometimes called Pomeron loops\cite{ploops}, or Fluctuations\cite{ms} or alternatively Saturation effects in the projectile wave function\cite{JIMWLK}.
It is the last characterization that is most suited to the nature of these lectures.
Somehow, although the physical underpinning of the low x (or high energy) evolution is the understanding how hadronic 
wave functions evolve with energy, the language of the wave function is not frequently used in this context. 
I believe that one can learn a lot by dealing directly with the wave function, since it contains all the nontrivial information there is to have, 
so my heart goes out to the wave function.
I have therefore decided for these lectures to adhere to its point of view on Life, the Universe and Everything, 
and as a small part of Everything, to show how to think about the evolution 
and derive the evolution equation from this perspective.

The high energy evolution in its latest incarnation became more or less synonymous with saturation.
Why is it interesting to study high energy evolution?
One obvious answer is of course that there is an experimental drive to understand the high energy regime. 
The resurgence of interest in this topic in the 90's was firmly triggered by the DIS data from HERA, which extended
experimental reach to $x\sim {\rm several} \times 10^{-5}$. The striking observation there was that the at low $x$ (high energy) the number of gluons in the proton shoots
up very fast (about 20 gluons at $Q^2=20$ GeV, $x=10^{-4}$). How and why does this number grow so fast? There is still no definitive answer to this question, in the sense 
that there are more than one fits to HERA data, some based on saturation\cite{satfits} and some not\cite{nonsatfits}, and it is strictly speaking not possible to differentiate between the physical
mechanisms involved on this basis. 
Nowadays this motivation is still there and is perhaps even stronger, since LHC is hopefully coming on line soon. 
There much larger numbers of gluons will be created in energetic p-p collisions.
Another motivation comes from RHIC. Although the energies there are not extremely large, due to the fact that the colliding objects are nuclei
the effects of the evolution are expected to appear much sooner. It may 
well be that large gluon densities are at work there. Some features of the data are at least qualitatively described by
the saturation models\cite{klm}, which are based in some aspects on low x evolution\cite{bkw}. And finally,  Pb-Pb collisions at LHC will lead to even higher gluon densities.

True as it may be, at the moment the high energy evolution and saturation  physics can not claim firmly that it is of phenomenological relevance.
For me personally the more appealing side of this field is that the physics is of fundamental theoretical interest.

We all know that QCD is nonperturbative in the Infrared. 
Masses, decay widths, elastic scattering cross sections can not be obtained without recourse to nonperturbative physics. 
But in the Ultraviolet the theory is perturbative. Quarks and gluons when probed on a short distance and time scales behave as if they are almost free and thus 
some physical processes can be calculated perturbatively. The prime example of this is DIS. Electron scatters off a nucleon with large (space like) momentum transfer $q^2=-Q^2$.
A simple picture of this process is given by the parton model.
\begin{figure}[ht]\epsfxsize=11.7cm
\centerline{\epsfbox{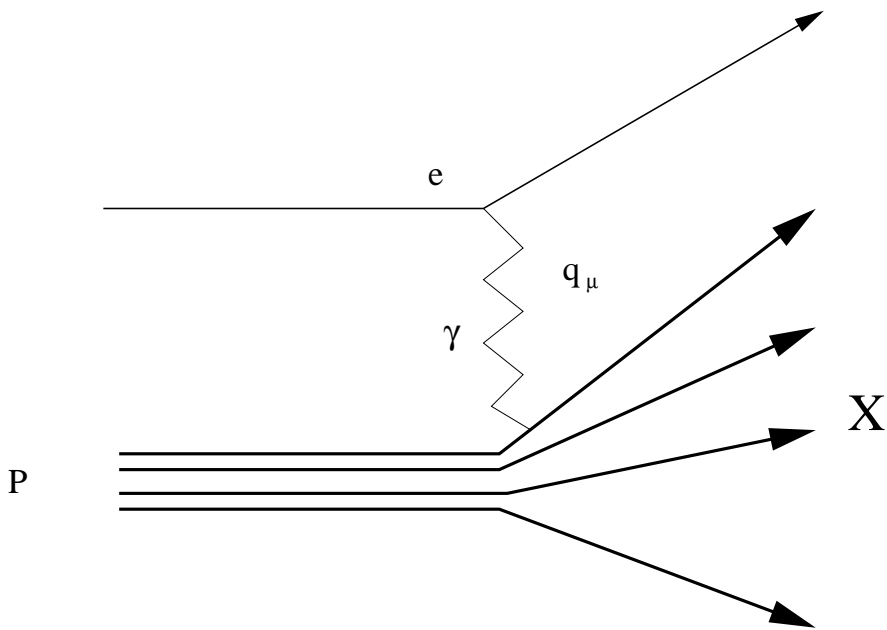}}
\caption{Deeply Inelastic Scattering
}\label{fig1}
\end{figure}
The proton comes in with momentum $p_\mu$ in the $z$ direction and is hit by a space like photon (emitted off the incoming electron). By measuring the outgoing electron one has access to
two kinematical variables, the energy transfer from the electron to the hadronic system, and the scattering angle of the electron. The two Lorentz scalar quantities
which encode this information are 
$Q^2$ and $x={Q^2\over 2P\cdot Q}$. The parton model is formulated in the infinite momentum frame of the proton, where the four momenta of the proton and the virtual
photon involved in the collision are
$$P_\mu(p)=(E_p=P,0,0,P),\ \ \ \ \ P\rightarrow\infty$$
$$q_\mu(\gamma)=(E_\gamma={Q^2\over 2xP}, Q[1+{Q^2\over 4x^2P^2}]^{1/2},0)$$ 
The energy of the colliding $\gamma-p$ system is $s=(Q_\mu+P_\mu)^2=Q^2({1\over x}-1)$ - at small $x$ obviously $s={Q^2\over x}$, hence small $x$ limit is the same as large 
energy limit in this case.

The physical picture of the parton model is simple. In the frame we have chosen, the photon is a very small probe of transverse size $\sim {1\over Q}$.
What does it see when it hits the proton? There is of course  a lot of stuff inside the proton, 
which is mainly soft muck. But a small probe does not see this - it whizzes through the muck unless it strikes something hard. That is to say, it only scatters off
the proton if it encounters another small object of roughly the same size as itself. The object in this case of course must be a quark, since only quarks are charged and interact with 
the photon. The scattering event itself therefore is a short distance process which also happens very quickly in time, with the relevant distance-time scale of order $1/Q$. 
At large $Q^2$ this is perturbative.

This short distance processes are described perturbatively, but such hard fluctuations over 
the soft background are very rare, thus the cross section is small. In other words there is only a small number of quarks in the proton that a hard probe can see.
The cross section for the $\gamma-p$ subprocess (forgetting about the $\gamma$ - emission vertex from the scattered electron) is easy to write down 
\begin{equation}
\sigma(Q^2)\propto {\alpha_{em}\over Q^2}N_q(Q^2,x)
\end{equation}
Here $1/Q^2$ is the size of the colliding objects, $\alpha_{em}$ is the probability that $\gamma$ scatters if it hits the quark, and $N_q$ is the average number of the 
hard quarks in the proton.
The parton model interpretation of the variable $x$ is the longitudinal momentum of the struck quark, or more precisely 
the fraction of total proton momentum that the struck quark happens to carry - hence its place of honor as a variable of which $N$ depends.
The source of this interpretation is the kinematics of the two body $\gamma-q$ collision. Think of the struck quark to be on shell, massless particle. Before the collision it carries
momentum $(k,0,k)$ since it moves in the same direction as the proton. After the collision its four momentum is $k'_\mu$
$$\left({Q^2\over 2xP}, Q,0\right)+(k,0,0,k)=k'_\mu$$
But  $(k')^2=0$, so $(k+{Q^2\over 2xP})^2-Q^2-k^2={kQ^2\over xP}-Q^2=0$. Thus $k=xP$.

The unknown quantity in the cross section is of course the number of quarks. 
We do not know it, since it is a nonperturbative quantity. It is some property of the proton
wave function, which develops over a long period of time and thus crucially depends on the nonperturbative physics. However one can certainly ask
within the perturbation theory how it changes, when we change the ``resolution scale'' $Q$, as long as $Q$ is large. 
If $Q$ increases, the change in the quark number is perturbative - it
changes simply due to very fast fluctuations. In a way, the old quarks and gluons ``split'' by perturbative radiation processes. As we increase the resolution of the 
probe, and also decrease the time duration of the interaction, the probe resolves more and more of these fast fluctuations in the wave function. This is described by 
the QCD DGLAP evolution equation.
\begin{figure}[ht]\epsfxsize=11.7cm
\centerline{\epsfbox{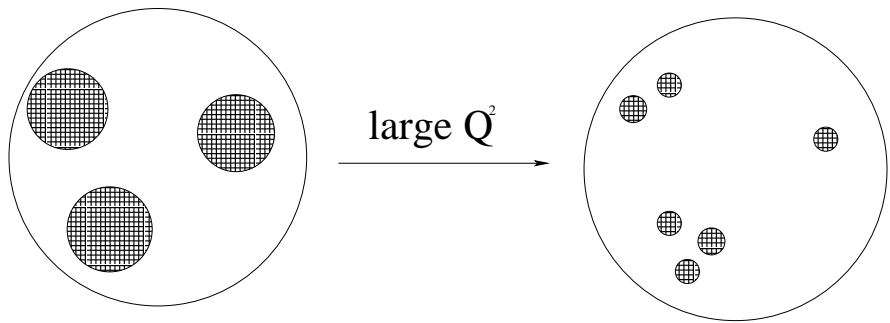}}
\caption{Cartoon of the DGLAP evolution. With improved transverse resolution the number of partons grows but the density decreases.
}\label{fig2}
\end{figure}
In this evolution the picture of the hadron basically does not change. There are a little bit more partons as we go to higher resolution 
(the growth is logarithmic in perturbative QCD), since when we look closer sometimes we see that what we thought was one parton, is in fact two sitting close together.
But the density of partons in the transverse space actually decreases, since the size of the resolved partons 
decreases much faster than their number increases.

But what happens if we increase the energy of the process keeping  the resolution fixed (decrease $x$ at fixed $Q^2$)? Again there will be a growth of 
number. The partons split and multiply by the same process of radiation of gluons. 
But the splitting now is predominantly in the longitudinal direction and on the same transverse scale. 
The partons split, but keep their transverse size. When looked ``face on'' 
the proton becomes denser and denser. 
\begin{figure}[ht]\epsfxsize=11.7cm
\centerline{\epsfbox{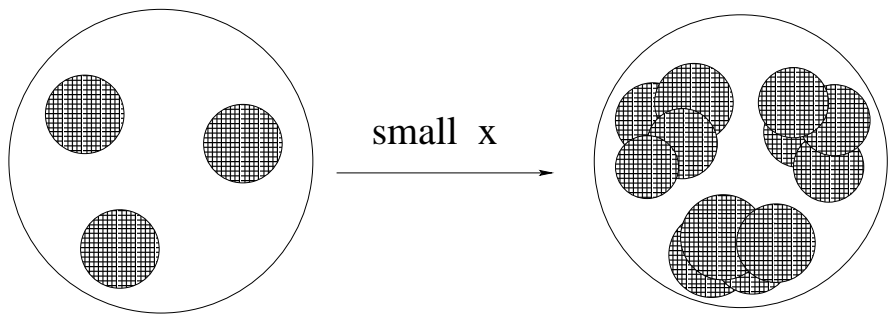}}
\caption{Cartoon of the low x evolution. As the energy increases the partons of fixed transverse size multiply. The density in the transverse plain grows.
}\label{fig3}
\end{figure}
 Again this is the effect of fluctuations. This time as it turns out (and as we will see later) the dominant contribution comes from gluons. So think about
it as if ``valence'' partons are surrounded by gluon cloud. The gluons in this cloud fluctuate very quickly, and are not seen by the probe, since their effect averages to 
zero during the time of the interaction. But increasing energy means that we are boosting our proton. Boost leads to Lorentz time dilation, so what used to
be fast becomes slow. The gluons ``freeze''. This Cold Gluon Cloud (CGC)\footnote{You have probably noticed that the abbreviation for Cold Gluon Cloud - CGC is the same as for Color Glass Condensate introduced in \cite{cgc}. This is in fact fortunate, since both refer to exactly the same thing. I personally prefer Cold Gluon Cloud  as it expresses better the essence of the relevant physics.} is indeed seen by the probe and is the main culprit for the growth of cross sections with energy.
The caveat is of course, that the photon does not see gluons directly, and so for DIS this effect comes into being slowly - only due to the fact that the sea
quark distributions grow due to the growth of gluon cloud. But for $p-p$ collisions for example one has direct gluon-gluon subprocesses, and those trigger directly
on the gluons in the proton.

Evolving far enough in energy will create a system where many gluons overlap in the transverse plane. The physics of the proton wave function becomes quite different -  
physics of a dense gluonic system, rather than that of a dilute small number of partons.
One has to be a little careful with the terminology - we are still looking at this system on relatively short time scales, on which the gluon cloud is static. 
Thus even though the system is dense, the physics is not that of thermodynamic equilibrium. Nevertheless the high density certainly has profound effect on
all kinds of observables.

What can we expect to see, if we collide two such dense "black" objects? Think of the limit when the whole face of the proton is covered by these overlapping frozen gluons. It is more
or less homogeneous on the low momentum scale, but is still of course grainy if we probe it at short enough distance. 
One can define the scale, which is proportional to the transverse
density of gluons. Let's call it the saturation momentum - $Q_s$. 
This scale separates the ``homogeneous'' regime from the ``grainy'' regime. For spatial resolution worse than $1/Q_s$ gluons strongly overlap in the transverse plane,
while at resolution better than $1/Q_s$ the gluonic system looks dilute. So $1/Q_s$ is the resolution at which the packing fraction of gluons in the transverse plane is 
about one half. 
As this is the characteristic transverse scale in the problem, we expect that after collision  
most final state gluons
will pick up this momentum $k_\perp\sim Q_s$. If $Q_s$ is large the physics at this scale is perturbative - we produce hard ``minijets''. These minijets will 
dominate the final state and so maybe we will even be able to describe the bulk of the cross section (total cross section?) in terms of these perturbative processes.
They are certainly not trivially calculable, as we have to learn how to deal with dense gluonic medium. Clearly naive perturbation theory in the sense I described above
does not work in this dense medium. But at least in principle we 
do not require the knowledge of soft physics of confinement and we expect with some effort to be able to figure the problem out.

Of course, the $k_\perp\sim Q_s$ is not the only type of physics that exists even in this regime. There are even more perturbative processes, like real hard 
jets. Those will stay
perturbative in the usual sense, although their multiplicities will be also affected, as presumably gluon distributions will change significantly in this regime.
There will also remain the soft physics. Proton has finite size, and clearly the density will grow in the center faster than on periphery. 
There will always be a peripheral region which 
is ``gray'' and not ``black'', that is where the analog of $Q_s$ is of order $\Lambda_{QCD}$. 
\begin{figure}[ht]\epsfxsize=11.7cm
\centerline{\epsfbox{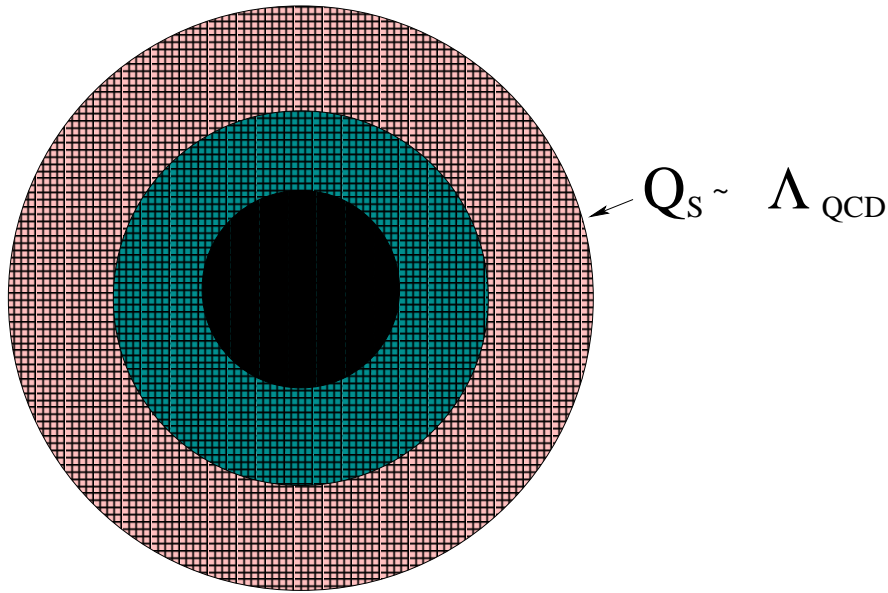}}
\caption{Schematic portrait of  a hadron at high energy. A black central region is surrounded by "gray" disc where the saturation momentum $Q_s$ is of order $\Lambda_{QCD}$.
}\label{fig4}
\end{figure}

We cannot hope to be able to describe the peripheral gray region in any kind of perturbative approach. Confinement rules there, and confinement therefore
will rule eventually for total and elastic cross sections. However it is conceivable that in some intermediate energy regime, the bulk of the proton is
already ``black'', but the peripheral region is still not too large. In this regime the total cross section can be dominated by the minijet processes, and 
might be describable perturbatively! It would be almost too good to be true if this did happen. This is a very long shot, and we would be extremely lucky. 
But in principle such a possibility exists and is very interesting to explore.

To summarize, some of the questions we may hope to address in the high energy - large density regime are

1. How does this blackness develop in the first place (Pomeron loops are presumably important)?

2. How does the density of the gluon cloud change with energy, and what is the density profile (transverse momentum dependence)?

3. What is the final state structure when colliding these ``black'' objects.

4. What can we say about the total cross section?

I will certainly not answer all of these questions in these lectures. In fact I will not even address most, and for those I will address only tentative answers will be provided.
Nevertheless a lot has been done and some has been understood in this area and there is ample room for further progress, which is enough of a motivation in itself.

So let's start climbing into the Cold Gluon Cloud.
We all fly a lot these days and know that going through a cloud could be often bumpy.  So fasten your seatbelts and let's go. 

The plan of these lectures is the following.

a) Derive the evolution equation (evolution in energy, x or rapidity) for the scattering of a dilute perturbative projectile on a dense target.
This is what is variably known as BK, JIMWLK, B-JIMWLK and B-JIMWLK$^2$ equation.

b) Discuss briefly what is known about the qualitative behavior of its solutions.

c). Discuss what kind of corrections have to be taken into account when the projectile itself becomes dense (Pomeron loops). Calculate some corrections
of this type and discuss a little bit what kind of physics we expect from them.

d). Prove an interesting duality relation which the complete evolution equation, including the Pomeron loops must satisfy.

\section{JIMWLK as we know it.}

Consider a small perturbative projectile with the wave function $|P>$ impinging from the left on a large dense target with the wave function $|T>$.
The target is dense, and therefore has large gauge fields. The projectile on the other hand is just a bunch of partons. In the following I will disregard quarks,
and will only consider gluons as the dynamical degrees of freedom.

We assume that both the target and the projectile move with close to velocity of light. Their longitudinal sizes are therefore Lorentz contracted in the lab frame into 
pancakes of almost zero width.
Thus all the projectile partons are localized around $x^-=0$, while the color fields of the target are 
at $x^+=0$. The only nontrivial dynamics therefore happens in the transverse plane.

We want first of all to derive an expression for the $S$-matrix for the collision of two such objects, and then see how it evolves with energy.
We assume that the initial energy is already quite large. 

\subsection{The forward amplitude}

First consider a single gluon in the projectile wave function. It has a large energy and therefore while traveling through the target it continues along the straight line.
Of course it could happen from time to time that it experiences hard scatterings and changes its direction of motion, but these events are rare and contribute 
very little to the total cross section. So the single most important 
thing that happens to the traveling gluon as a result of the interaction is that its color degree of freedom rotates.
\begin{equation}
|x,a\rangle\equiv a^{a\dagger}_i(x)|0\rangle\rightarrow S^{ab}(x)|x,b\rangle
\end{equation}
The $S$-matrix operator in this approximation is block diagonal. It is diagonal in the space of transverse coordinates, particle number etc, and is nontrivial only in the
space of color indices.
The matrix $S^{ab}$ must be unitary, since the $S$-matrix is a unitary operator.
The forward scattering amplitude is defined as the overlap of the $|in\rangle$ and $|out\rangle$ states, and thus is given by 
the unitary matrix we have just defined
\begin{equation}
_{\rm in}\langle {\rm gluon},b|{\rm gluon},a\rangle_{\rm out}=\langle 0| a^a_i(x)\hat Sa^{\dagger b}_i(x)|0\rangle=S^{ab}(x)
\end{equation}

Now in fact one can express the matrix $S$ in terms of the target fields through which the gluon is propagating. The target of course is a quantum object,
and therefore the fields in it fluctuate. However the propagation happens very fast, as we assume that the energy of the projectile is large. 
Therefore during the propagation the projectile really sees a fixed field configuration.
Since we are considering the projectile to consist of partons, it is convenient to use the light cone gauge, in which the wave function of the projectile has
partonic description, namely $A^+=0$. In this gauge naturally the largest component of the field in the target is $A^-$. The target moves with large rapidity in the
 negative $x_3$ direction.
Recall that under the Lorentz boost in the negative $x_3$ direction the components of vector potential transform as
\begin{eqnarray}
&&A^-(x^-,x^+,x_\perp)\rightarrow e^{Y} A^-(e^{Y} x^+, e^{-Y}x^-,x_\perp)\nonumber\\
&&A^+(x^-,x^+,x_\perp)\rightarrow e^{-Y} A^+(e^{Y} x^+, e^{-Y}x^-,x_\perp)\nonumber\\
&&A_i(x^-,x^+,x_\perp)\rightarrow  A_i(e^{Y} x^+, e^{-Y}x^-,x_\perp)
\end{eqnarray}
where $Y$ is the boost parameter.
Thus a large $A^-$ field component is generated in the target when accelerating it. The $S$ matrix of a fast particle interacting with the
$A^-$ field is given by the eikonal formula
\begin{equation}
S^{ab}(x)=\left[P\exp\left\{ig\int dx^+ \alpha^c(x,x^+)T^c\right\}\right]^{ab}
\label{alpha}
\end{equation}
where $g$ is the strong interaction coupling constant, $\alpha=A^-$ (change of notation for historical reasons only), $T^a_{bc}=if^{abc}$ is 
the $SU(N)$ generator in the adjoint representation and $P$ denotes the path ordering along the trajectory of the propagating gluon..
Now, the fields $\alpha(x)$ are of course operators in the target Hilbert space. So to calculate the $S$-matrix for the process, we have to 
average over the target wave function.
\begin{equation}
s(1\ {\rm  gluon})=\langle T|S(\alpha)|T\rangle=\int D\alpha(x)S[\alpha]W_T[\alpha]
\label{walpha}
\end{equation}
We have introduced here the weight function $W_T[\alpha]$ which should be thought of as the square of the target wave function and thus gives the probability density
to find a particular configuration of $\alpha$ in the wave function of the target.
There is a subtlety involved here, and the identification of $W$ with the square of the wave function is not entirely correct\cite{kl}. But it is good enough for all we are going to do it these lectures.
We can also redefine $W$ in such a way that the basic variable becomes $S$.
So
\begin{equation}
s(1\ {\rm  gluon})=\int DS(x)S(x)W_T[S]
\label{ws}
\end{equation}
with the understanding that $W$ is normalized as the probability density
\begin{equation}
\int DSW_T[S]=1
\end{equation}
The functionals $W_T$ in equations (\ref{walpha}) and (\ref{ws}) strictly speaking differ by the Jacobean of the transformation (\ref{alpha}), but for the simplicity of 
notation I will not indicate this explicitly.

Consider now an arbitrary projectile whose wave function in the gluon Fock space can be written as
\begin{equation}
|P\rangle=\Psi[a^{\dagger a}_i(x)]|0\rangle
\label{wf}
\end{equation}
The gluon creation operator $a^\dagger$ depends on the transverse coordinate, and also on the longitudinal momentum $k^+$. 
We assume that all the gluons are energetic, and therefore all gluon operators which enter eq.(\ref{wf}) have longitudinal momenta above some cutoff $\Lambda$. 
We will refer to these degrees of freedom as "valence". The longitudinal momenta of the gluons are not explicitly written down in eq.(\ref{wf}) because
we assume that as long as those gluons are energetic enough they all scatter eikonally, and thus their amplitude does not depend on the exact value of the momentum.

Now, since the interaction with the target takes a very short time, all gluons scatter independently of each other. Thus 
the outgoing state simply has every gluon creation operator multiplied by the single gluons scattering matrix at the corresponding transverse position.
\begin{equation}
\hat S|P\rangle=\Psi[S^{ab}(x)a^{\dagger b}_i(x)]|0\rangle
\end{equation}

Given this, the formal expression for the $S$ - matrix of the projectile can be written as
\begin{equation}
\Sigma[S]\equiv \langle P|\hat S|P\rangle=\langle 0|\Psi^*[a^{ a}_i(x)]
\Psi[S_{ab}(x)a^{\dagger b}_i(x)]|0\rangle
\end{equation}

 The forward scattering amplitude for the process is thus given by
\begin{equation}
s({\rm P\ on \ T})=\langle \Sigma\rangle_T\,=\,\int \,dS\,\, \Sigma[S]\,\,W_T[S]
\label{average}
\end{equation}

Now given these engredients, we ask how does the forward amplitude change as we make more energy available for the collision.

\subsection{What of boost?}

We can increase the center of mass energy of the colliding system by either boosting the target or the projectile. Of course it does not matter, since the S-matrix is 
Lorentz scalar. The JIMWLK equation, which is what we set out to derive, is the equation for the evolution of the target probability density function $W_T$. However it is
more convenient to derive it by considering boosting the projectile, and then imposing the requirement of Lorentz invariance of the amplitude.

So let's think what happens if we boost the projectile. In what we did so far, we assumed that we know the projectile wave function at longitudinal momenta greater 
than some cutoff $\Lambda$. We also tacitly assumed that the number of particles below this cutoff is small. Otherwise they would also contribute to the S-matrix.
So even though we do not know exactly how the wave function looks for $k^+<\Lambda$, we are not particularly worried. However once we start to be interested in boosting,
this becomes a very important question. The point is that in QCD boosting a state changes the number of gluons in it. Let's pause and ask ourselves how does this happen?

Let's say we boost the state with some rapidity $y$. We know that one effect of this is to scale the longitudinal momenta of all the particles present in the wave function 
by 
\begin{equation}
k^+\rightarrow e^yk^+
\end{equation}
So if we had particles sitting above some cutoff $\Lambda$ and the phase space below the cutoff was small, then the cutoff moves and the phase space increases.
\begin{figure}[ht]\epsfxsize=11.7cm
\centerline{\epsfbox{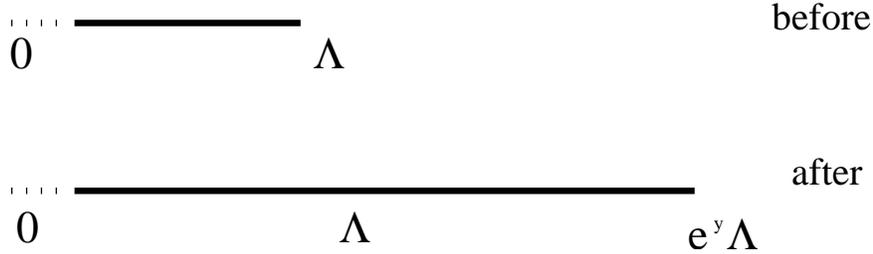}}
\caption{Growth of the longitudinal phase space under boost.
}\label{fig5}
\end{figure}
The boost drags particles from under the cutoff into the ``observable'' range of momenta. The ``unknown'' phase space is stretched:
\begin{equation}
P.S= \int^{\Lambda_2}{dk^+\over k^+}\rightarrow \int^{e^y\Lambda_2}{dk^+\over k^+}=P.S.+y 
\end{equation}

By itself this is not a big problem. If the number of particles does not change, then it does not matter much if their longitudinal momenta have
increased - in our calculation the scattering amplitude does not depend on the value of the longitudinal momentum.
The trouble is that in QCD things are more complicated. 
Remember: gluons (or photons) are transverse modes of the color field (electromagnetic field). In principle the color field also has a longitudinal 
component - Coulomb field, which is non dynamical. However 
boost does not preserve the longitudinal nature of the field, and thus transforms the Coulomb field into live gluons.

Think as an example of a static charge. It is accompanied by the Coulomb electric field
\begin{equation}
E^i(x)={g\over 4\pi}{r^i\over |r|^3}
\end{equation}
This is purely longitudinal. Now boost it with a very large rapidity. It's a little bit too cumbersome to write down explicit expressions for finite boost, so 
let us formally write it for infinite boost. We know that what you get is the electric field which is shrunk to a $\delta$-function in the longitudinal direction, 
while since the total flux is not changed the transverse components decrease much slower at large transverse distances.
\begin{equation}
E^i(x)={g\over 4\pi}{r^i\over |r|^3}\rightarrow E^{i}={g\over 2\pi}{x_\perp^i\over x_\perp^2}\delta(x^-)
\end{equation}
This is the so called Weizsaker-Williams (WW) field.
It obviously has transverse components since 
$\partial_3E^i\ne 0$. So by boosting a benign Coulomb field we have created live gluons/photons. We can actually calculate how many 
we have created.
We will do it more formally in the next lecture, but for now we only have to remember how the gluon creation and annihilation operators are related to the electric field
\begin{equation}
E^i(k)=i\sqrt{ \omega(k)}[a_i(k)-a_i^\dagger(k)]=i\sqrt {k^+}[a_i(k)-a_i^\dagger(k)]
\end{equation}
where the last line assumes large longitudinal momentum.
For the Weizsaker-Williams field
\begin{equation}
E^i(k)=ig{k^i_\perp\over k_\perp^2}
\end{equation}
So we find
\begin{equation}
a(k)\sim g{1\over \sqrt {k^+}}{k^i_\perp\over k_\perp^2}
\end{equation}
and the total number of gluons at a given transverse momentum
\begin{equation}
N(k_\perp)=\int dk^+\langle a^\dagger_i(k)a_i(k)\rangle={\alpha_s\over k_\perp^2}\int {dk^+\over k^+}={\alpha_s\over k_\perp^2}y
\end{equation}
That's quite impressive: we have created the number of particles proportional to the total longitudinal phase space opened up by the boost.

If we boost by a large $y$, we create lots of gluons out of the Coulomb field. And we do have Coulomb field: 
we have valence gluons in our wave function. They have color charges, and
therefore by fiat create Coulomb field, and this will turn into new gluons after boost;
the new gluons also have color charge, and so they carry their own Coulomb field and SO IT GOES ON.

Now, the Coulomb field (or the zero mode, or whatever you want to call it) is really everything there is below our cutoff $\Lambda$. 
To know how many gluons will be created by the boost, we therefore need to peak under the cutoff.

The lesson is that in order to find the evolution of $S$ we need to know the wave function below $\Lambda$ to greater accuracy than simply saying there are 
almost no gluons there. 
This is our next task. W are now going to address the question: given that we know the wave function above $\Lambda$, 
how do we find  it below $\Lambda$?

\subsection{The Light Cone Hamiltonian.}
Our aim now is to find the projectile wave function at low longitudinal momenta. First of all we have to set up the Hamiltonian calculation. Since our projectile moves very 
fast, the formalism that is most suited for it is the Light Cone Quantization. This has been used for many years now in the context of the parton model, but also more 
recently to study the spectrum of QCD. A comprehensive review of the formalism is given in \cite{brodsky}.

The essence of this setup is that one quantizes not on the equal time surface $t=0$ and dynamically evolves states in time, but rather on the surface $x^+=0$.
The evolution in $x^+$ (which is the analog of the time variable) is generated by the operator $P^-=H-P_3$. This is done in the light cone gauge $A^+=0$.
Yet another component of the vector potential is eliminated, since one of the Maxwell's equations becomes a constraint on the light cone:
\begin{equation}
\partial^+\partial^+A^-+\partial^+\partial_iA^i=0
\end{equation}
which is solved by
\begin{equation}
A^-=-{1\over \partial^+}\partial_iA^i
\end{equation}
This is very convenient, since only the transverse field components are left as dynamical degrees of freedom and all non dynamical fields
are already eliminated. The curios feature of the Light Cone Quantization (LCQ) is that it has only half of the number of degrees of freedom compared to 
the equal time quantization. Only positive longitudinal momentum components of the 
field create particles, while negative longitudinal components annihilate them. This is understandable, since we want to describe the situation where
all the particles move right with large energy. A state like this can not by definition contain particles with negative momentum, and thus
negative momentum cannot correspond to particle creation, but only to annihilation. The mathematical manifestation of this is that the fields $A^i$
contain both, canonical coordinates and canonical momenta and thus 
do not commute at 
different values of the longitudinal coordinate $x^-$.
When all is said and done, and when the quantization is performed properly,
the LCQ formalism amounts to the following.

a). The ``time'' coordinate is $x^+$, the ``space'' coordinates are $x^-,\ \ x_i$.

b). The basic degrees of freedom are the transverse components of vector potential with the commutation relation
  
\begin{equation}
 \left[ A_a^i(x^-,{\bf x}), A_b^j(y^-,{\bf y}) \right]
  = {i\over 2}\epsilon(x^--y^-)\, \delta_{ab}\, \delta^{ij}\,
             \delta^{(2)}({\bf x} - {\bf y})\, ,
\end{equation}
where $\epsilon (x)$ is the antisymmetric step function $\epsilon (x)={1\over 2}{\rm sign} (x)$.
In momentum space the commutator reads
\begin{equation}
  \left[ A_a^i(k^+,{\bf x}), A_b^j(p^+,{\bf y}) \right]
  = \frac{2\pi}{2k^+}\, \delta(k^+ +p^+)\, \delta_{ab}\, \delta^{ij}\,
             \delta^{(2)}({\bf x} - {\bf y})\, ,
\end{equation}             
The field $A$ can be represented in the standard way in terms of the creation and annihilation operators
\begin{eqnarray}
  &&A^i_a(x^-,{\bf x}_\perp) = \int_0^\infty
  \frac{dk^+}{2\pi} \frac{1}{\sqrt{2k^+}}
  \Bigg\lbrace a^i_a(k^+,{\bf x})\, e^{-ik^+x^-}
  + a_a^{i\, \dagger}(k^+,{\bf x})\, e^{ik^+x^-}
  \Bigg\rbrace\, ,
  \nonumber \\
 &&\left[ a_a^i(k^+,{\bf x}), a_b^{j\, \dagger}(p^+,{\bf y})
   \right] = (2\pi)\, \delta_{ab}\, \delta^{ij}\, \delta(k^+ -p^+)\,
             \delta^{(2)}({\bf x} - {\bf y})\, .
\end{eqnarray}

c) The time evolution of the wave functions in this Hilbert space is generated by the following light cone Hamiltonian (LCH)(we switched notations from $P^-$ to $H$ for simplicity):
\begin{eqnarray}
  H =  \frac{1}{2\pi}\int_{k^+>0}   \left[ \frac{1}{2} \Pi_a^-(k^+,{\bf x})\, \Pi_a^-(-k^+,{\bf x})
      + \frac{1}{4} G_a^{ij}(k^+,{\bf x})\, G_a^{ij}(-k^+,{\bf x})
      \right]
\end{eqnarray}
where the electric and magnetic pieces are
\begin{eqnarray}
  \Pi_a^-(x^-,{\bf x}) &=& \frac{-1}{\partial^+}
       \left( D^i\partial^+A^i\right)_a(x^-,{\bf x})
       \\
  G_a^{ij}(x^-,{\bf x}) &=&
  \partial^i A^j_a(x^-,{\bf x}) -  \partial^j A^i_a(x^-,{\bf x})
     - g f^{abc}\, A_b^{i}(x^-,{\bf x})\, A_c^{j}(x^-,{\bf x}) \nonumber
\end{eqnarray}

d) The physical wave functions are subject to the Gauss' law constraint
\begin{equation}
  C_a({\bf x}) = \int dx^-\, (D^i\partial^+A^i)_a(x^-,{\bf x}) = 0\, .
\label{constr}
\end{equation}
This constraint is the reflection of the fact that $A^+=0$ is not a complete gauge fixing. It still leaves the residual 
freedom to perform gauge transformations
with the gauge function which does not depend on $x^-$. Hence the integral over $x^-$ in the constraint eq.(\ref{constr}).

The standard perturbation theory can be formulated straightforwardly in this framework. First of all, one notices that even though the theory is interacting, 
the exact ground state of this Hamiltonian is the Fock space vacuum of the operators $a(k)$. This is easy to understand. Since the Hamiltonian commutes with
the total momentum operator in the longitudinal direction, $P^+$, when written in momentum space each and every term in this Hamiltonian has to have vanishing total 
longitudinal momentum. But this immediately means that there can be no terms which contain only creation operators of the type $(a^\dagger)^n$, since all 
creation operators have
positive longitudinal momentum! Thus each term must contain at least one annihilation operator, and therefore the Fock space vacuum of $a$ is also the vacuum of $H$.

At any rate the free Hamiltonian is
\begin{equation}
H_0=\int_{k^+>0}{dk^+\over 2\pi}{d^2k_\perp\over(2\pi)^2}{k_\perp^2\over 2k^+}a^{\dagger a}_i(k^+,k_\perp)a^{a}_i(k^+,k_\perp)
\end{equation}
and this defines the standard free dispersion relation
\begin{equation}
2k^-k^+-k_\perp^2=0,\ \ \ \ \ \ \ \ \ \  k^-={k_\perp^2\over 2k^+}
\end{equation}

We are however not interested in the vacuum state. We want to examine the state which contains some particles at valence momenta $k^+>\Lambda$, and in particular to
find how this state looks like at momenta below $\Lambda$.
There is a very natural framework in which to approach this problem, namely the Born-Oppenheimer approximation. The free dispersion relation $k^-={k_\perp\over 2k^+}$
tells us that the frequencies
of the high momentum modes are small while those of small momentum modes are large. Thus in the framework of the light cone evolution, the high longitudinal momentum 
modes are slow degrees of freedom, and we can study the low momentum mode wave function by momentarily freezing the high $k^+$ part. This of course is very much in line
with the picture of the Gluon Cloud which cools and freezes under the boost transformation.

The most important characteristic of the high momentum component of the state is its colour charge density
\begin{equation}
\rho^{a}(x_\perp)=\int_{k^+>\Lambda}a^{\dagger b}_i(k^+,x_\perp)T^a_{bc}a^{c}_i(k^+,x_\perp)
\end{equation}
since the eikonal $S$-matrix depends only on the colour charge density 
at every point in the transverse plane. 

So the strategy is the following. We will forget about the self interaction of the high $k^+$ modes in the Hamiltonian. We will keep the self interaction of low $k^+$
modes as is. And additionally we will keep the eikonal interaction between the fast and slow modes, namely the interaction between the low $k^+$ modes and the
color charge density generated by the valence modes.

One can question the applicability of the Born-Oppenheimer approximation, since the spectrum of frequencies in the problem is actually continuous, so there is no big 
frequency gap between what we call ``fast'' and what we call ``slow''. Fortunately however the $k^+<\Lambda$ bin is dominated by modes with momentum much smaller 
than $\Lambda$. The relevant corrections are proportional to the total phase space available in the low $k^+$ bin, and that goes as $y=\ln (\Lambda/\Lambda_0)$. This
means that the momenta that dominate the integrals are of order $k^+\sim\sqrt {\Lambda\Lambda_0}\ll\Lambda$.

I will not further justify this approximation here, but it turns out to be precisely the same eikonal approximation which we used to define the scattering matrix.

All said and done the coupling between the valence and the soft modes appears in the Hamiltonian through the modified electric contribution
\begin{eqnarray}
 && \Pi_a^-(k^+,{\bf x}) = \frac{-1}{-i(k^++i\epsilon)}
  \partial^i \Pi_a^i(k^+,{\bf x})
  + \frac{1}{-i(k^++i\epsilon)} \rho_a({\bf x})
 \\
  &&+ g \frac{1}{-i(k^++i\epsilon)} f_{abc}
  \int_{|p^+|<\Lambda} \frac{dp^+}{2\pi}
  A_b^i(k^+-p^+,{\bf x})\, (-ip^+)\, A_c^i(p^+,{\bf x})\, \nonumber  .
\end{eqnarray}
Now, although $\rho$ is an operator on the valence Hilbert space, as far as the soft fields are concerned, it is a $c$-number function. Also we will assume, 
consistently with our discussion so far that the number of valence gluons is parametrically of order one, and thus also
\begin{equation}
\rho^a\sim O(1)
\end{equation}
This makes life much easier, as the interaction term is multiplied by the QCD coupling constant. So we can apply perturbation theory. Let us therefore
calculate the lowest order perturbative correction to the vacuum wave function of the soft gluons in the presence of valence charge density. 
As already explained, to zeroth order the vacuum of the LCH is simply the Fock space vacuum of $a$. To first order
the correction to the vacuum is given by
\begin{equation}
|{\bf vacuum}\rangle=|0\rangle+\Sigma_i|i\rangle{\langle i|\delta H|0\rangle\over E_i-E_0}
\end{equation} 
where the summation is over the eigenstates of the zeroth order Hamiltonian, and $E_i$ denotes the zeroth order energy of the respective eigenstate.
What makes the calculation simple, is that to first order only the interaction term that involves the valence charge gives a contribution. The reason is that, as we have 
noted before the Fock space vacuum is actually annihilated by all terms of the LCH, as they contain at least one annihilation operator. This is not the case however for the 
soft-valence coupling
\begin{equation}
\delta H=\int {dk^+\over 2\pi}{d^2k_\perp\over (2\pi)^2}{gk_i\over |k^{+}|^{3/2}}\left[a^{\dagger a}_i(k^+,k_\perp)+a^a_i(k^+,k_\perp)\right]\rho^a(-k_\perp)
\end{equation}
This Hamiltonian creates only one particle state from the vacuum:
\begin{equation}
\langle k_\perp,k^+,a,i|\delta H|0\rangle={gk_i\over |k^{+}|^{3/2}}\rho^a(-k_\perp)
\end{equation}
With $E_i$ determined by the free dispersion relation
we can write the soft gluon vacuum state to first order in the coupling as 
\begin{equation}
 |{\rm vacuum}\rangle=\left\{1+
i\int d^2xb_i^a(x)\int_{k^+<\Lambda}{dk^+\over \pi^{1/2}| k^+|^{1/2}} a^{\dagger a}_i(k^+, x)\right\}|0\rangle
 \label{vacuum}
\end{equation}
where the ``classical'' field $b_i$ is precisely the Weizsaker-Williams field of the colour charge density $\rho^a$
\begin{equation}
b^a_i(x)={g\over 2\pi}\int d^2y{(x-y)_i\over (x-y)^2}\,\rho^a(y)\,.
\end{equation}
Actually one has to be a little bit more careful. We have found the correct wave function, but it is not properly normalized. 
The normalization strictly speaking is the 
second order correction in $g$, but it has to be restored if we want to have unitary evolution. This is easily done, by simply subtracting 
from the coefficient of the Fock vacuum term the total probability to have a one gluon state in the wave function.

\begin{eqnarray}
 |{\rm vacuum}\rangle&=&\left\{\left[1-{1\over 2 \pi}\int_{k^+<\Lambda}{dk^+\over k^+}\int d^2x(b_i^a(x)b_i^a(x))\right]\right.\nonumber\\
&+&i\int d^2xb_i^a(x)\int_{k<\Lambda}\left.
{dk^+\over \pi^{1/2} | k^+|^{1/2}} a^{\dagger a}_i(k^+, x)\right\}|0\rangle
 \label{wfun}
\end{eqnarray} 
In fact, as will become obvious later, this is the only $O(\alpha_s)$ correction to the wave function which gives contribution of $O(\alpha_s)$ 
to the evolution equation.

So we found what we were looking for - the Weizsacker-Williams field hiding behind the cutoff.
Boosting the state we will pull out this field from underneath the cutoff and into the wide open world. 
Thus starting at the initial rapidity $Y_0$ with some valence state with momenta above $\Lambda$, call it
$|P\rangle_{Y_0}=|P,k^+>\Lambda\rangle$ 
and boosting it by the boost parameter $\delta Y$ we obtain
\begin{eqnarray}
&&|P\rangle_{Y_0+\delta Y}=\left\{\left[1-{1\over 2 \pi}\int_{\Lambda}^{(1+\delta Y)\Lambda}{dk^+\over k^+}\int d^2x(b_i^a(x)b_i^a(x))\right]\right.\\
&&+i\int d^2xb_i^a(x)
\int^{(1+\delta Y)\Lambda}_{\Lambda}\left.
{dk^+\over \pi^{1/2} | k^+|^{1/2}} a^{\dagger a}_i(k^+, x)\right\}|P,k^+>(1+\delta Y)\Lambda\rangle\nonumber
\end{eqnarray}
We have boosted all the existing gluons to higher momentum, plus have produced new gluons all over the newly opened phase space. It is straightforward to see, that
the number of newly produced gluons is simply proportional to the phase space and to the intensity of the Weizsacker-Williams field
\begin{equation}
\int_{\Lambda}^{(1+\delta Y)\Lambda}dk^+\langle a^{\dagger a}_i(k^+,x_\perp)a^{a}_i(k^+,x_\perp)\rangle={\delta Y\over \pi}b^a_i(x)b^a_i(x)
\end{equation}

\subsection{Forty two}

Given that we know now the evolution of the wave function, we can find the evolution of 
the $S$-matrix. We go a little back to recall the eikonal expression for the $S$-matrix

\begin{eqnarray}
s({\rm P\  on\  T})&=&\langle \Sigma\rangle_T\,=\,\int \,dS\,\, \Sigma[S]\,\,W_T[S]\\
\Sigma[S]&\equiv& \langle P|\hat S|P\rangle=\langle 0|\Psi^*[a^{ a}_i(x)]
\Psi[S_{ab}(x)a^{\dagger b}_i(x)]|0\rangle\nonumber
\label{reminder}
\end{eqnarray}

The evolved wave function still scatters eikonally.
\begin{eqnarray}
&&\hat S|P\rangle_{Y_0+\delta Y}=
\left\{\left[1-{1\over 2 \pi}\int_{\Lambda}^{(1+\delta Y)\Lambda}{dk^+\over k^+}\int d^2x(b_i^a[S\rho,x]b_i^a[S\rho,x])\right]\right.\\
&&\left.+i\int d^2xb_i^a[S\rho, x]
\int^{(1+\delta Y)\Lambda}_{\Lambda}
{dk^+\over \pi^{1/2} | k^+|^{1/2}} S^{ab}(x)a^{\dagger b}_i(k^+, x)\right\}\times\nonumber\\
&&\ \ \ \ \ \ \ \ \ \ \ \ \ \ \ \ \ \ \ \ \ \ \ \hat S|P,k^+>(1+\delta Y)\Lambda\rangle\nonumber
\end{eqnarray}
Here the charge density that creates the Weizsacker-Williams field in the outgoing wave function is also rotated by the same one gluon scattering matrix
\begin{equation}
b^a_i[x,S\rho]={g\over 2\pi}\int d^2y{(x-y)_i\over (x-y)^2}\,S^{ab}\rho^b(y)\,.
\end{equation}

Since the initial projectile wave function does not contain soft gluons, the calculation of the averages of the soft gluon operators is straightforward.
The S-matrix of the boosted projectile is therefore (to first order in $\delta Y$):
\begin{eqnarray}
&&\Sigma_{Y+\delta Y}[S]\,=\,\Sigma_{Y}[S]\,\\&&-\,{\delta Y\over 2\pi}\,
\langle 0|\,\Psi^*[a^{ a}_i]\,\int d^2z\,
\left[b_i^a([z,\rho])\,b_i^a[z,\rho]\,+\,b_i^a[z,S\rho]\,b_i^a[z,S\rho]\right.\nonumber\\
&&\left.-
2\,b_i^a[z,\rho]\,S^{ba}(z)\,b_i^b[z,S\rho]\right]\,
\Psi[S_{ab}a^{\dagger b}_i]\,|0\rangle\nonumber\label{deltasigma}
\end{eqnarray}

Our aim is actually to represent this expression in terms of functional derivatives of $\Sigma$ with respect to $S$. This involves some algebra, but it is quite cute.
First we note that the charge density operator, when acting on the wave function simply rotates the gluon creation operator at the transverse position 
in question. We can therefore write 
\begin{eqnarray}
\rho^a(x)\Psi[Sa^{\dagger}_i]|0\rangle&=&T^a_{bc}a^{\dagger b}_i(x){\delta\over\delta a^{\dagger c}_i(x)}
\Psi[Sa^{\dagger}_i]|0\rangle\nonumber\\
&=&-{\rm tr} \left\{S(x)T^{a}{\delta\over \delta S^\dagger(x)}\right\}\Psi[Sa^{\dagger}_i]|0\rangle\ \ ;
\end{eqnarray}
The second equality is simply the reflection of the fact that in an expression of the form $Sa$, the left multiplication of $a$ is equivalent to 
the right multiplication of $S$.
So the charge density operator acts like the right rotation on the matrix S
\begin{equation}
\rho^a(x)\{S^{bc}(x)\}=-[S(x)T^a]^{bc}
\end{equation}
Since our wave function is assumed to be dilute (does not contain more than one gluon at each point in space) we can immediately write
\begin{eqnarray}
&&\rho^b(y)\rho^a(x)\Psi[Sa^{\dagger}_i]|0\rangle=\\
&&\ \ \ \ \ \ \ {\rm tr} \left\{S(x)T^{a}{\delta\over \delta S^\dagger(x)}\right\}{\rm tr} 
\left\{S(y)T^{b}{\delta\over \delta S^\dagger(y)}\right\}\Psi[Sa^{\dagger}_i]|0\rangle \nonumber
\end{eqnarray}
It is also easy to see that the operator $S\rho$ acts as the left rotation:
\begin{eqnarray}
 &&\left[S(x)\rho(x)\right]^a\Psi[Sa^{\dagger}_i]|0\rangle =
 -{\rm tr} \left\{T^{a}S(x){\delta\over \delta S^\dagger(x)}\right\}\Psi[Sa^{\dagger}_i]|0\rangle\ \ ;\nonumber\\
 &&\rho^b(y) \left[S(x)\rho(x)\right]^a\Psi[Sa^{\dagger}_i]|0\rangle \nonumber\\
 &&\ \ \ \ \ \ \ \ \ =
 {\rm tr} \left\{T^{a}S(x){\delta\over \delta S^\dagger(x)}\right\}{\rm tr} \left\{S(y)T^{b}{\delta\over \delta S^\dagger(y)}\right\}\Psi[Sa^{\dagger}_i]|0\rangle\; \nonumber\\ && etc.
\label{sun}
\end{eqnarray}
We can thus immediately write
\begin{equation}
{\delta\over\delta Y}\,\Sigma[S]\,\,=\,\,\chi[S,{\delta\over\delta S}]\,\,\Sigma[S]
\end{equation}
where the kernel $\chi$ is given by 
\begin{eqnarray}
\chi^{JIMWLK}&=&-\,\frac{\alpha_s}{2\,\pi^2}
\int_{x,y,z}{(z-x)_i(z-y)_i\over (z-x)^2(z-y)^2}\times \,\\
&&\left[ J^a_L(x)J^a_L(y)+J^a_R(x)J^a_R(y)
-\,2\,J^a_L(y)J^b_R(x)S^{ba}(z)\right]\nonumber
\label{smallevolution}
\end{eqnarray}
with the right- and left rotation operators are defined as
\begin{eqnarray}
J_R^a(x)&=&-{\rm tr} \left\{S(x)T^{a}{\delta\over \delta S^\dagger(x)}\right\};\ \  J_L^a(x)=
-{\rm tr} \left\{T^{a}S(x){\delta\over \delta S^\dagger(x)}\right\};\nonumber\\
J_L^a(x)\,\,&=&\,\,[S(x)\,J_R(x)]^a,
\end{eqnarray}

This is the JIMWLK equation. Actually, those who have seen JIMWLK before will probably not quite recognize it in this formula. For starters, JIMWLK is
usually written as the evolution equation for the target probability density function $W_T$. This is easy to do using Lorentz invariance of the forward scattering amplitude.
Recall that the forward scattering amplitude is given by the 
equation (\ref{reminder}). Let us restore the rapidity labels on various quantities. The $S$ matrix is defined at 
rapidity of interest $Y$. But we can think of this rapidity as distributed between the target and the projectile. 
Let's say that the target has been boosted to rapidity $Y_0$, 
and the projectile to rapidity $Y-Y_0$, so that eq.(\ref{reminder}) properly reads
\begin{equation}
s_Y({\rm P\  on \  T})=\langle \Sigma^{Y-Y_0}\rangle_{T(Y_0)}\,=\,\int \,dS\,\, \Sigma^{Y-Y_0}[S]\,\,W^{Y_0}_T[S]\\
\end{equation}
Lorentz invariance requires that $s$ does not depend on $Y_0$. Requiring that the derivative of $s$ with respect to $Y_0$ vanishes, we find immediately
\begin{eqnarray}\label{smallevolutionW}
{\delta\over\delta Y}\,W_T[S]\,\,=\,\,\chi^{JIMWLK}[S,{\delta \over\delta S}]\,\,W_T[S]\,.
\end{eqnarray}
This is still not the form in which JIMWLK is usually written down. To get it into that form, recall that the eikonal
expression of the single gluon $S$-matrix is given by
\begin{equation}
S(x)=P\exp\left\{i\int_{-\infty}^{Y} dx^- \alpha^a(x,x^-)T^a\right\}
\end{equation}
with $P$ denoting the path ordering. When acting on any function of $S$, the derivative with respect 
to $\alpha(x,Y)$ is identical to the action of $J_R(x)$:
\begin{eqnarray}
{\delta\over\delta\alpha^e(x,Y)}F[S]&=&{\delta S^{ab}(x)\over\delta\alpha^e(x,Y)}
{\delta\over\delta S^{ab}(x)}F[S]\nonumber\\&=&i[S(x)T^e]^{ab}{\delta\over\delta S^{ab}(x)}F[S]=-iJ_R^e(x)F[S]
\label{equiv}
\end{eqnarray}
Thus we can write eq.(\ref{smallevolutionW}) in the familiar form
\begin{eqnarray}
&&\chi^{JIMWLK}=\frac{\alpha_s}{2\pi^2}\int_{x,y,z} {(z-x)_i(z-y)_i
\over (z-x)^2(z-y)^2}\times\\
&& {\delta\over \delta \alpha^a(x,Y)}
\left[1+S^\dagger(x)S(y)-S^\dagger(x)S(z)-S^\dagger(z)S(y)\right]^{ab}
{\delta\over \delta \alpha^b(y,Y)}\nonumber
\label{JIMWLK}
\end{eqnarray}

Let me finish this lecture with some remarks regarding the JIMWLK equation.
First, as is clear from the above derivation, the introduction of the additional coordinate $x^-$ and the path ordering in eq.(\ref{alpha}) is not really necessary. It 
merely serves to mimic the 
action of $iJ_R$ on a function of $S$ as in eq.(\ref{equiv}). 
In the framework of JIMWLK equation it has no independent meaning, since none of the physical observables depend on $x^-$. JIMWLK lives happily without
the additional coordinate and is defined exclusively in terms of the unitary matrix $S$ and its functional derivative $\delta/\delta S$.
I have only written it here in this form for those who have seen it before and are used to this particular notation.
My personal preference is clearly not to introduce unnecessary variables, so I prefer the much simpler form eq.(\ref{smallevolution}). 

The second remark is about the applicability of this equation. We have derived it here assuming the projectile is small and deriving the evolution of the 
projectile wave function. On the other hand the original 
derivation was given directly for the target wave function, assuming explicitly that the target is dense, that is that the average of the single gluon $S$
matrix is close to zero \cite{JIMWLK}. How come we get the same equations under seemingly unrelated conditions? The answer is that these two conditions - diluteness
of the projectile and dense nature of the target are actually inextricably related by the form of the $S$-matrix eq.(\ref{reminder}). Mathematically 
diluteness of the projectile is simply the statement that the projectile averaged $S$-matrix $\Sigma$ has only low order terms in the Taylor expansion in powers of $S$.
On the other hand, if the target is dense it means that the probability density functional $W_T$ is strongly peaked around $S=0$. If that is the 
case, clearly eq.(\ref{reminder}) will again be dominated by the first few terms in the Taylor expansion of $\Sigma$. Thus the diluteness of projectile, and the dense nature
of the target in eq.(\ref{reminder}) go hand in hand.

Finally the last remark is related to the operatorial nature of $S$. As I have mentioned before, the single gluon scattering matrix is in fact an operator in the 
Hilbert space of the target. Nevertheless we have allowed ourselves to treat it as a $c$-number field neglecting possible commutators. The reason we can do this is
because the target is dense. What this means, is that the charge density of the target is large. Parametrically JIMWLK equation corresponds to
the target charge densities $\rho\sim {1\over\alpha_s}$. In this regime the commutator $[\rho^a,\rho^b]=iT_{ab}^c\rho^c$ is 
suppressed  by $\alpha_s$ relative to  the product of $\rho$'s and can therefore be neglected. This makes $\alpha$'s also 
classical and together with them the matrices $S$ themselves.

\section{JIMWLK and friends}

Let me now discuss some conspicuous limits of the JIMWLK equation and some properties of its solutions.

\subsection{The dipole model}
Since its introduction by Al Mueller \cite{dipole}, the dipole degrees of freedom have become very popular in the discussion of high energy scattering.
The formal justification for this model is the large $N_c$ limit. In this limit, in any diagram a gluon line can be represented as a double line
of a fictitious quark and antiquark. It turns out that the diagrams in which these fictitious quarks and antiquarks form singlet states - dipoles, are leading
in the large $N_c$ limit. The vertex of emission of the WW gluon then is simply represented as the splitting of one dipole into two.
\begin{figure}[ht]\epsfxsize=11.7cm
\centerline{\epsfbox{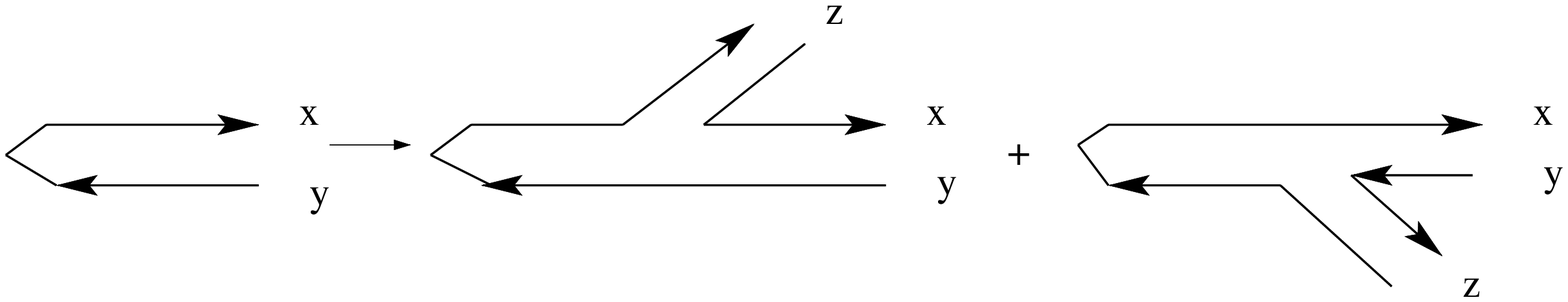}}
\caption{In the large $N_c$ limit emission of one gluon off a dipole is equivalent to the splitting of the  dipole into two.
}\label{fig6}
\end{figure}

The simplicity of the dipole picture, is that there is no color degree of freedom anymore, so that the $S$-matrix of a dipole is a single number rather than an $SU(N)$ 
matrix. Since the color was the only degree of freedom in which the eikonal $S$-matrix was not diagonal, the $S$ matrix in the dipole model is in fact diagonal.
We can derive the dipole model limit of the JIMWLK equation in a very simple way. The $S$ matrix for a dipole is related to the single gluon $S$-matrix as:
\begin{equation}
s(x,y)={1\over N}{\rm Tr}(S^\dagger_F(x)S_F(y))
\end{equation}
where now $S_F$ is the fundamental representation of the single gluon scattering matrix (which is the scattering matrix of the fictitious quark which constitutes the dipole).
We now assume that the projectile and the target consist entirely of dipoles, that is that the probability 
density functional $W$ and the projectile scattering matrix $\Sigma$ depends only on this particular combination $W(s)$, $\Sigma(s)$.
It is then straightforward to see that the action of the left and right color rotations in the leading $N_c$ approximation
reduce to the following

\begin{eqnarray}\label{dipole}
J_R^a(y)\Sigma[s]\,&=&\,\frac{1}{N_c}\,\,{\rm tr}[S^\dagger(u)S(v)\tau^a]\,\,\,
[\delta(v-y)-\delta(u-y)]\,\,{\delta \Sigma\over \delta s(u,v)}\nonumber\\
J_L^a(y)\Sigma[s]\,&=&\,\frac{1}{N_c}\,{\rm tr}[S^\dagger(u)\tau^aS(v)]\,\,\,
[\delta(v-y)-\delta(u-y)]\,\,{\delta \Sigma\over \delta s(u,v)}\nonumber\\
J_R^a(x)J_R^a(y)\Sigma[s]\,&=&\,\frac{1}{N_c}\,
{\rm tr}[S^\dagger(u)S(v)\tau^a\tau^a]\,\,\,
[\delta(v-y)-\delta(u-y)]\,\nonumber\\
&\times&[\delta(v-x)-\delta(u-x)]\,\,
{\delta \Sigma\over \delta s(u,v)}+\,O\left(\frac{1}{N^2_c}\right)\nonumber\\
J_L^a(x)J_L^a(y)\Sigma[s]\,&=&\,\frac{1}{N_c}\,\,{\rm tr}[S^\dagger(u)S(v)\tau^a\tau^a]\,\,\,
[\delta(v-y)-\delta(u-y)]\,\nonumber\\&\times&[\delta(v-x)-\delta(u-x)]\,\,
{\delta \Sigma\over \delta s(u,v)}+\,O\left(\frac{1}{N^2_c}\right)\nonumber\\
J_L^b(x)J_R^a(y)\Sigma[s]
\,&=&\,\frac{1}{N_c}\,{\rm tr}[S^\dagger(u)\tau^bS(v)\tau^a]\,\,
[\delta(v-y)-\delta(u-y)]\,\nonumber\\
&\times&[\delta(v-x)-\delta(u-x)]\,\,{\delta \Sigma\over \delta s(u,v)}+\,O\left(\frac{1}{N^2_c}\right)
\end{eqnarray}
Here $\tau^a$ are the generators of $SU(N)$ group in the fundamental representation. 
We have also dropped the subscript $F$ for brevity, but all the matrices in eq.(\ref{dipole}) are in the fundamental representation.
To simplify this further we need to use the completeness relation
\begin{equation}
\tau^a_{\alpha\beta}\tau^a_{\gamma\delta}=\frac{1}{2}
\left[\delta_{\alpha\delta}\delta_{\beta\gamma}-
{1\over N_c}\delta_{\alpha\beta}\delta_{\gamma\delta}\right ]
\end{equation}
and the identity that relates adjoint and fundamental representations of a unitary matrix
\begin{equation}\label{AF}
S_A^{ab}(z)\,=\,2\,Tr\,\left[ \tau^a\,S_F(z)\,\tau^b\,S_F(z)^\dagger\right]
\end{equation}
Working out the color algebra we find that the evolution operator $\chi^{JIMWLK}$ in the large $N_c$ limit reduces to
\begin{eqnarray}\label{dec}
&&\chi^{JIMWLK}[S|_{N_c\rightarrow\infty}]\,=\,\chi_{dipole}[s]=
\,\frac{ \bar{\alpha}_s}{2\,\pi}\,
\int_{x,y,z}\frac{(x-y)^2}{(x-z)^2\,(z\,-\,y)^2}\,\,\nonumber\\
&&\times\left[\,-s(x,\,y)\,+\,
\,s(x, z)\,s(y,z)\,\,\right]
\frac{\delta}{\delta s(x, y)} 
\end{eqnarray}
The large $N_c$ limit  evolution
\begin{equation} \label{W1}
\,\frac{\partial \,\Sigma[s]}{
\partial \,Y}\,\,= \,\, \chi_{dipole}[s]\,\,\Sigma[s];
\,\,\,\,
\,\frac{\partial \,W[s]}{
\partial \,Y}\,\,= \,\,\chi_{dipole}[s]\,\,W[s]\,.
\end{equation}
can be also written as
\begin{equation}\label{S03}
\frac{d\,\hat s(x,y)}{d\,Y}\,=\,-\,\bar \alpha_s \int d^2z\,
\frac{(x-y)^2}{(x-z)^2\,(y-z)^2}\,\,\left[\hat s(x,y)\,-\,\hat s(x,z)\,\hat
s(y,z)\right ]
\end{equation}
where the operator $\hat s$ is introduced to indicate that this equation generates the hierarchy of equations for the target averages of the 
powers of $s(x,y)$ - multi dipole scattering amplitudes. In particular the single dipole amplitude satisfies
\begin{eqnarray}\label{S04}
\frac{d\,\langle s(x,y)\rangle}{d\,Y}\,&=&\,-\,\bar \alpha_s \int d^2z\,
\frac{(x-y)^2}{(x-z)^2\,(y-z)^2}\,\,\nonumber\\
&\times&\left[\langle s(x,y)\rangle\,-\,\langle s(x,z)
s(y,z)\rangle\right ]
\end{eqnarray}
The forward amplitude is related to the scattering probability by
\begin{equation}
s(x,y)=1-N(x,y)
\end{equation}
Thus eq.(\ref{S04}) is the equation for the evolution of the single dipole scattering probability on the target
\begin{eqnarray}\label{S05}
&&\frac{d\,\langle N(x,y)\rangle}{d\,Y}\,=\,\,\bar \alpha_s \int d^2z\,
\frac{(x-y)^2}{(x-z)^2\,(y-z)^2}\,\,\\
&&\times\left[\langle N(x,z)+N(y,z)-N(x,y)-\,\ N(x,z)N(y,z)\rangle\right ]\nonumber
\end{eqnarray}
In this way of writing it is obvious that the meaning of the last term is simply the probability of simultaneous scattering of both
dipoles off the target.
If we assume that the areas of the target on which the dipoles scatter are uncorrelated, we can forget about the target averages and write down a closed equation for the 
single dipole probability
\begin{eqnarray}\label{S06}
&&\frac{d\,N(x,y)}{d\,Y}\,=\,\,\bar \alpha_s \int d^2z\,
\frac{(x-y)^2}{(x-z)^2\,(y-z)^2}\,\,\\
&&\times\left[ N(x,z)+N(y,z)-N(x,y)-\,\ N(x,z)N(y,z)\right ]\nonumber
\end{eqnarray}
This is the Kovchegov equation \cite{kovchegov}. Note that it is derived assuming factorization of the dipole scattering probabilities. This is best valid when the 
target is dense and when the correlation length of the target fields is small, smaller than the distance between the two dipoles. 
Strictly speaking the factorization of the nonlinear term is 
not valid when the target is far from saturation, that is $N<1$. However in this regime the nonlinear term is
not important anyway, and so it came to be that the numerical results obtained studying the Kovchegov equation \cite{bknum} so far 
are very close to the results obtained from the whole JIMWLK hierarchy \cite{jimwlknum}.

\subsection{The BFKL limit}
Omitting the nonlinear term $N^2$ in (\ref{S04})we recover the BFKL equation. 
It is of course always nice to see a familiar face, however 
on the negative side, the BFKL equation is known to be problematic. It violates unitary in more than one way. That is to say it violates both the unitarity of 
the scattering probability (the probability can not exceed one) and the unitarity of the total cross section (the cross section can not grow faster than the Froissart bound 
$\sigma^{total}<{\pi\over m^2} y^2$ in any massive theory).

The first violation is due to the fact that the solution of the BFKL equation for $N$ grows without bound. BFKL is a linear equation and thus if the kernel has a 
positive eigenvalue clearly the probability grows. Indeed the BFKL solution at asymptotically large rapidities is\cite{salam}
\begin{equation}
N(x,b)\sim  \exp\{\omega Y-\ln {16b^2\over x_0 x} -{\ln^2
{16b^2\over x_0 x}\over a^2Y}\})
\end{equation}
Here $\omega=4\ln 2 N_c\alpha_s/ \pi$, $a^2=14\zeta(3)N_c\alpha_s/\pi$ and $\zeta(n)$ being the Riemann zeta function,  
with $x$ -the size of the dipole, and $b$ -the impact parameter.
This unitarity problem is cured by JIMWLK. This is easiest to see on the level of Kovchegov equation. Clearly in the presence of the nonlinear 
term the point $N(x)=1$ is the fixed point of the equation. If one starts with small $N$, it cannot grow beyond unity.

In fact, the evolution works in the way quite consistent with our initial expectations about the generation of a typical intrinsic momentum scale.
In particular, at every rapidity $Y$, there is a dipole size $x_s(Y)$ such that for the dipoles of larger sizes the target is black 
(the scattering probability is close to unity), while for smaller size dipoles it is basically transparent. The inverse of this scale is the saturation momentum $Q_s(Y)$.
From the point of view of the target evolution, this is the scale on which the target is grainy - you probe it at smaller distances, you are liable to see
very little since the fields are correlated and the phase rotation of $q$ and $\bar q$ are the same, while at larger separation the phases rotate randomly and the 
scattering is almost unavoidable. I will discuss properties of $Q_s$ some more in a short while.

However there is also another problem, and that is that the BFKL probability amplitude grows large at large impact parameters. In particular in the BFKL 
solution the probability is of order one at impact parameters all the way to 
\begin{equation}
b\sim e^{\kappa\omega Y}
\end{equation}
with $\kappa$ somewhat smaller than one, but not significantly so. As long as $N$ is smaller than one, the nonlinear term in the Kovchegov equation is 
not effective, thus it is clear that in this peripheral region the solution of the nonlinear equation is similar to that of BFKL.
The total cross section is given by
\begin{equation}
\sigma^{total}=\int d^2bN(b)\sim e^{2\kappa\omega Y}
\end{equation}
This blatantly violates the Froissart bound. The qualitative picture is that of the growth of the dipole due to the long range nature of the WW fields \cite{froissart}. 
Since perturbative gluon is massless, this growth in the transverse space is very fast, which leads to the violation of the Froissart bound.
\begin{figure}[ht]\epsfxsize=11.7cm
\centerline{\epsfbox{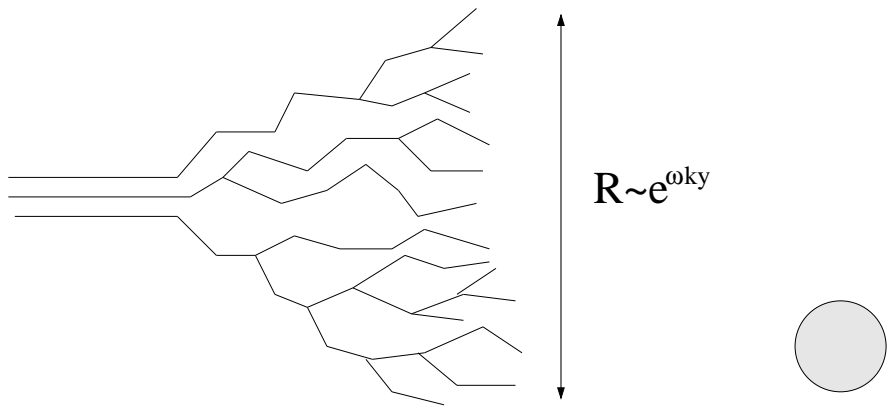}}
\caption{Violation of the Froissart bound by the perturbative nonlinear QCD evolution. The transverse size of the hadron grows exponentially with rapidity due to emission of massless gluons. If initially the projectile hadron missed the target, it will hit it after a long evolution due to this growth of transverse size.
}\label{fig7}
\end{figure}
To solve this problem one has to understand the real Infrared scale of QCD - how is it that gluons do not travel farther than $\Lambda_{QCD}$ from
their parents. This problem is outside the scope of B-JIMWLK$^2$, and so there is no real wonder that it stays unresolved.

\subsection{The saturation momentum and the geometric scaling.}
Not all is doom and gloom though. Even though the total cross section at very high energies is nonpertubative, it does not mean one cannot extract useful physics 
from JIMWLK. Recall, the main motivation to consider the physics of saturation is to study the ``minijet physics'' around and about the saturation scale. This can be
done, since the main mechanism of the generation of $Q_s$ is local and is unaffected by the nonperturbative Coulomb tails.
The properties of the solutions of the Kovchegov equation have been studied very extensively \cite{bknum}, but the main properties can be understood in a very simple way \cite{scaling}. Basically 
the solutions to Kovchegov equation is the same as that of BFKL for transverse sizes for which $N<1$, while for sizes where the BFKL solution violates
unitarity, the BK solution is simply unity.
A good approximation to the BFKL solution at large energy is (one have to integrate the $b$ dependence in the formula above)
\begin{equation}
N(x)=\alpha_s^2 (x^2\Lambda^2)^{1-\lambda_0}\exp\{\omega Y\}
\label{Nas}
\end{equation}
where $\Lambda$ is the scale that characterizes the target at initial rapidity (not the longitudinal cutoff of the previous lecture!), and $\lambda_0=0.38$ is a pure number.
The crossover scale $Q_s$ is determined by equating this expression to a number of order unity. One finds
\begin{equation}
x_s^2\Lambda^2=\exp\{-{\omega\over 1-\lambda_0}Y\}
\end{equation}
which determines the saturation momentum
\begin{equation}
Q_s^2={1\over x_s^2}=\Lambda^2\exp\{{\omega\over 1-\lambda_0}Y\}
\end{equation}
The interesting feature of this solution is that if now we rewrite the amplitude in terms of $Q_s(Y)$ (still in the regime where it is small)
we obtain
\begin{equation}
N(x)=(x^2Q_s^2(Y))^{1-\lambda_0}
\end{equation}
Thus there is no separate dependence on $x$ and $Y$, but only on one particular combination of the two quantities. This phenomenon came to be known as geometric scaling.
Of course, generally one expects such a behavior for any saturated system at transverse momenta around $Q_s$. Here however the surprising thing is, that 
due to the fact that the amplitude is a simple power, the scaling persists to values of momentum much larger than $Q_s$.
A simple estimate of the values of $x$ where the scaling breaks down goes as follows. There are corrections to the BFKL solutions at small values of $x$.
The leading correction is the appearance in the exponent of the same factor ${\ln^2(x^2\Lambda^2)\over a^2Y}$
This factor becomes comparable with the terms that we kept
in eq.(\ref{Nas}) when ${\ln^2(x^2\Lambda^2)\over a^2Y}=\omega Y$. Reexpressing this in terms of the saturation momentum, we find that
the corrections are small all the way to 
\begin{equation}
x^2\Lambda^2= \left({\Lambda\over Q_s^2}\right)^{n}, \ \ \ \ \ n\sim 2
\end{equation}
So it turns out that in the BFKL solution geometric scaling persists to dipole sizes very much smaller than $1/Q_s$. Although this is a very interesting property, one 
has to be careful when trying to assess its phenomenological relevance. 
It does not appear to be a universal prediction of saturation but a specific property of the solution of the BFKL equation.
Therefore, even though the DIS data clearly exhibits such scaling \cite{golec} for small values of $x$, its description in terms of the solution of JIMWLK equation is only as good as the
BFKL fit to DIS data \cite{royon}.

\section{What's wrong with JIMWLK?}

What we have accomplished so far is the derivation of the evolution equation in the leading order in QCD coupling constant, in the eikonal approximation and 
assuming dilute projectile and dense target. So quite a number of restrictions and any one of them when lifted may change results dramatically. 

In particular,
we know that higher order corrections to BFKL equation are very large. And this remains the main threat even if the nonlinearities are accounted for. Will the 
infrared part of emission kill the whole idea of perturbatively probing saturation? We don't know, but we also can do nothing about it at this point, except try and analyze these corrections \cite{nexttoleading}.

What about the eikonal approximation? How long can we rely on it? Here we pretty much know that it will break down eventually, but we can estimate when it will happen.
Assume that we start with a dipole with initially large energy $E_0$. The saturation momentum of the target is what governs the recoil effects which are neglected 
in the eikonal approximation. As the saturation momentum becomes of the order of the initial energy, 
\begin{equation}
Q_s(Y)=E_0
\end{equation}
the eikonal approximation should be abandoned. 
Even assuming that the saturation momentum does grow exponentially (which is certainly not the case when next to leading corrections are taken into account \cite{nexttoleading}) 
this condition gives
\begin{equation}
Y={1\over\alpha_s}\ln {E_0\over\Lambda_{QCD}}
\end{equation}
This is pretty far away, so the question is really academic. There is plenty of rapidity to go before we reach this limit, and so we can safely use eikonal approximation
for years to come.

Now what about the nature of the projectile-target system? This question is the focus of very intense investigation in the last year. It goes under various names, 
like Pomeron loops, fluctuations or wave function saturation effects. And this is what I want to talk about in the rest of these lectures.

So why should we bother? Well, JIMWLK is strictly applicable only when the target is dense. Amazingly and surprisingly, its formal low density limit
reduces to the BFKL equation, which is the correct equation at low densities. But what we really want to understand is the transition between the dilute and the dense regime.
And in this region, for sure things are missing from JIMWLK. 
Let us be faithful to the projectile wave function picture, and look at the question from this point of view. Here we have made the approximation that the density is small.
For example our expression for the WW field was first order in the color charge density
\begin{equation}
b_i\propto{\partial_i\over\partial^2}\rho
\end{equation}
This approximation is responsible for the very fast exponential growth of the projectile density (and so the saturation momentum). Remember, we have found that the 
number of gluons is proportional to $b^2$? We can also easily calculate the change in the color charge density itself
\begin{equation}
\rho^a=\rho^a_{\rm valence}+a_{\rm soft}^\dagger T^aa_{\rm soft}
\end{equation}
In our wave function
\begin{equation}
\langle\delta\rho^a(x)\delta\rho^a(y)\rangle\approx\int_{u,v}{(x-u)\cdot(y-u)\over(x-v)^2(y-v)^2}\rho^a(u)\rho^a(v)
\end{equation}
Roughly speaking  we have a linear homogeneous equation for the evolution of the 
projectile density (we can be more precise and see that the charge density correlator actually satisfies the BFKL equation)
\begin{equation}
{\delta\rho^2\over\delta Y}=K\rho^2
\end{equation}
The solution to which is clearly an exponential in Y
\begin{equation}
\rho^2(Y)=\exp\{KY\}\rho^2_0
\end{equation}
The exponential growth of the projectile density is the origin of the exponential growth of the saturation momentum $Q_s$. The physics of this is very simple - the number of
gluons created by evolution is proportional to the number of gluons (color charges) already existing in the state. That's because every gluon in the wave function
emits independently of the presence of other gluons. But of course as a result of such evolution the projectile wave function itself 
becomes dense, at which point independent emission approximation is not valid anymore. In fact two things will happen when $\rho$ becomes large, 
and they combine to give an interesting picture.
First,there will be nonlinear effects in the emission. Simplistically, we can think of this by just assuming that the emissions are happening not in the empty space, but in
the presence of a non vanishing background field (the WW field itself). The emission vertex then
should be modified to something like
\begin{equation}
\delta b_i\propto{D_i\over D^2}\rho
\end{equation}
where $D$ is the covariant derivative in the $b$-background, $D=\partial-b$. This has the effect that when the field $b$ is 
large (or better to say for momenta smaller than the scale determined by $b$) the emitted field is not proportional to $\rho$ anymore, but is simply of order unity
\begin{equation}
\delta b={b\over b^2}\rho\sim {\rm independent \ of} \ \rho
\end{equation}
There is also another effect, that I will call ``bleaching of color''. When the projectile is dilute, all gluons are emitted into points in space where there were no other gluons previously. This means that the charge density at these points monotonically grows. But suppose we are now in the dense regime, so that a gluon is 
emitted with large probability into a point where there already exists a large color charge. 
\begin{figure}[ht]\epsfxsize=11.7cm
\centerline{\epsfbox{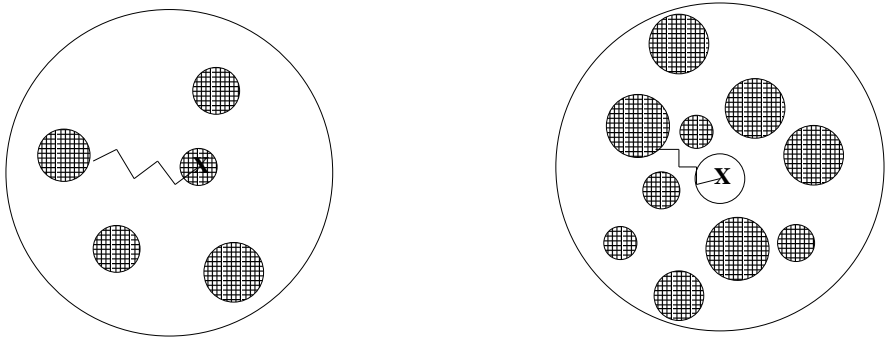}}
\caption{Bleaching of colour. If a gluon is emitted into the transverse position where there already is a gluon, the total color at this point may become zero.
}\label{fig8}
\end{figure}

What happens to the color at this point? Well, the color Casimir there is as likely to decrease as it is likely to increase.  The color charge density will random walk!
In fact, coupled with our expectation that the emission probability of the gluon is constant and independent of density, we expect genuine random walk process.
\begin{equation}
\rho^2(Y)\sim\rho^2_0+KY
\end{equation}
Much, much slower than exponential! This clearly must have a large effect on physics. We do not understand yet very well what quantities 
will change significantly and which will not. For example $Q_s$ may not change too much, since it is mainly sensitive to the range of momenta where $\rho$ is 
borderline small. But it certainly will effect the structure of final states at momenta smaller than $Q_s$ and may well also affect the properties of the gluon
distribution above $Q_s$. 

At any rate looks like fun physics to investigate and understand.

Incidentally, the bleaching of color, I believe is the same as Pomeron loops. The gluon is emitted, but is not actually seen, as its color is bleached by
the gluon emitted in the next rapidity step. So effectively this is a loop.

To include the saturation effects in the projectile wave function is therefore the same as including Pomeron loops in the high energy evolution. Do we know how to do that?
Not so far, although there have been a lot of progress in this direction recently. I will tell you a little about the results obtained so far, 
and you will hear more about it from Dionysis' talk later on\cite{dionysis}. 

\subsection{The simpleminded fix.}

Let's go back to our LCH and see if we can do something when the color charge is not too small.
\begin{eqnarray}
&&  H =  \frac{1}{2\pi}\int_{k^+>0}   \left[ \frac{1}{2} \Pi_a^-(k^+,{\bf x})\, \Pi_a^-(-k^+,{\bf x})
      + \frac{1}{4} G_a^{ij}(k^+,{\bf x})\, G_a^{ij}(-k^+,{\bf x})
      \right]\nonumber\\
&& \Pi_a^-(k^+,{\bf x}) = \frac{-1}{-i(k^++i\epsilon)}
  \partial^i \Pi_a^i(k^+,{\bf x})
  + \frac{1}{-i(k^++i\epsilon)} \rho_a({\bf x}) \\
  && + g \frac{1}{-i(k^++i\epsilon)} f_{abc}
  \int_{|p^+|<\Lambda} \frac{dp^+}{2\pi}
  A_b^i(k^+-p^+,{\bf x})\, (-ip^+)\, A_c^i(p^+,{\bf x})\,\nonumber\\ 
  &&G_a^{ij}(x^-,{\bf x}) =
  \partial^i A^j_a(x^-,{\bf x}) -  \partial^j A^i_a(x^-,{\bf x})
     - g f^{abc}\, A_b^i(x^-,{\bf x})\, A_c^j(x^-,{\bf x})\, ,\nonumber
\end{eqnarray}

As before, the Hamiltonian has a linear term, except that now $\rho$ is not small. We usually solve such problems by shifting the field. It's not quite straightforward
now since $\rho$ is itself an operator (albeit on a different Hilbert space). However let's forget about it for a moment and treat it like a $c$-number.
Then we can get rid of the annoying linear term by the unitary transformation
\begin{equation}
A\rightarrow C^\dagger AC
\end{equation}
with the cloud operator
\begin{equation}
 C=  \exp \Bigg\lbrace i\int_{p^+>\Lambda} d y_\perp {dp^+\over2\pi|p^+|^{1/2}}
       \left[a^{\dagger a}_i(p^+, y_\perp)+a^{a}_i(p^+, y_\perp)
\right]\, b_a^i( y_\perp)
                 \Bigg \rbrace
\end{equation}
The ``classical" field $b$ now solves the full nonlinear Yang-Mills equation:
\begin{eqnarray}
&& \partial_i\,  b_a^i( x_\perp) = \rho_a(x _\perp)\, ,  \nonumber \\
&& \partial_i\,  b_a^j( x_\perp) -
  \partial_j\,  b_a^i(x_\perp) - g f_{abc}\, b^i_b( x_\perp)\, b^j_c( x_\perp) = 0\, .
\end{eqnarray}

This corresponds to the classical solution for the gauge potential of the form
\begin{equation}
A_i^a(x)=\theta(x^-)b_i^a(x_\perp)
\end{equation} 
of the Yang-Mills equations with the source term 
\begin{equation}
J^{+a}(x^-,x_\perp)=\delta(x^-)\rho^a(x_\perp)
\end{equation}
Note that the field $b$ is a two dimensional pure gauge, so it can be written as
\begin{equation}
b^a_i={i\over g}{\rm tr}[\tau^aU^\dagger\partial_i U]
\end{equation}
It's easy to see that this works, because with this classical field, both $\Pi^-$ and $G^{ij}$ classically vanish, and thus classically this certainly minimizes 
the Hamiltonian. It is also clear that for large $\rho$ this eliminates the ``large'' contribution to the energy which can be as big as $O(1/\alpha_s)$.
In this approximation the vacuum wave function for the soft gluons is a coherent state given by the action of the cloud operator $C$ on the Fock space vacuum. 
In fact, if we take the extra rapidity interval to be small (small boost), we can expand the exponential in $C$ to second order in $b$. Note that the expansion parameter here is not $\alpha_s$ anymore, as the field is $O(1/g)$, but the width of the rapidity interval. We can then repeat the derivation of the boosted state and find
\begin{eqnarray}
&& |\Psi(Y+\delta Y)\rangle=\left\{\left[1-{1\over 2\pi}
\delta Y\int d^2z(b_i^a(z,[\rho])b_i^a(z),[\rho])\right]\right.\\
&&\left.+i\int d^2z \,
b_i^a(z,[\rho])\int_{\Lambda}^{(1+\delta Y)\Lambda}{dk^+\over
\sqrt\pi |k^+|^{1/2}} a^{\dagger a}_i(k^+, z)\right\}|\Psi(Y)\rangle\nonumber
\end{eqnarray}
The only difference with the expression we derived previously is that the classical field now is related to the charge density via a nonlinear equation.
Again we can continue our previous derivation to find that the kernel of the evolution kernel now becomes:
\begin{eqnarray}
\chi^{JIMWLK+}=&-&\,\frac{\alpha_s}{2\,\pi^2}
\int_{z}\left[b^a_i(z,J_R)b^a_i(z,J_R)+b^a_i(z,J_L)b^a_i(z,J_L)\right.\\
&&\left.-
2b^a_i(z,J_L)b^a_i(z,J_R)S^{ba}(z)\right]
\label{smallevolution+}
\end{eqnarray}
Where as before the right- and left rotation operators are 
\begin{eqnarray}
&&J_R^a(x)=-{\rm tr} \left\{S(x)T^{a}{\delta\over \delta S^\dagger(x)}\right\}; \ J_L^a(x)=
-{\rm tr} \left\{T^{a}S(x){\delta\over \delta S^\dagger(x)}\right\};\nonumber\\
&&J_L^a(x)\,\,=\,\,[S(x)\,J_R(x)]^a,
\end{eqnarray}
This time $b$ is a fairly complicated differential operator which satisfies the equation
\begin{eqnarray}
 && \partial_i\,  b_a^i(x_\perp) = J_R(x_\perp)\, ,
  \nonumber \\
  &&\partial_i\,  b_a^j( x_\perp) -
  \partial_j\,  b_a^i(x_\perp) - g f_{abc}\, b^i_b( x_\perp)\,
                                    b^j_c(x_\perp) =0\, .
\end{eqnarray}

This expression has been derived by Misha Lublinsky and myself \cite{kl},\cite{kl1}, and also recently  rederived by Hatta et.al. in an intriguing paper using the formalism of effective action \cite{smith}.
It would be nice of course, if this was the whole story, but unfortunately it is not. If one does things more carefully  it turns out that
the shift by the classical field mixes positive and negative longitudinal momenta. This means that the vacuum of the transformed Hamiltonian is not the Fock space vacuum 
any more, but rather there is a Bogoliubov transformation which transforms one into the other. Thus the soft vacuum is not simply $C|0\rangle$ and more work has to be 
done to find it. These corrections, at least on the face of it are not parametrically suppressed relative to the coherent transformation. 

So we do not quite know the evolution kernel yet. 
There are however some interesting hints that we can get from these expressions. The propagation of the gluon carries an eikonal factor $S$. On the other hand the classical
field of the projectile can be expressed in terms of matrices $U$, which also can be written as eikonal-like factors, but this time in terms of the projectile charge density,
which as we have seen becomes a derivative with respect to the phase of the eikonal factor. This is suggestive of some kind of duality, just like duality between $x$ and $p$
for a simple harmonic oscillator. Indeed, as it turns out, it is possible to prove that 
Lorentz invariance plus the condition that the projectile and target wave functions evolve in the same way (projectile-target democracy) requires that the
evolution be self dual \cite{duality}. In the last lecture I discuss in more detail the derivation of this self duality property.

\subsection{The self duality of high energy evolution}

We start by considering a general expression for the $S$-matrix of a projectile with the wave function 
$|P\rangle$ on a target with the wave function $|T\rangle$, where the total rapidity of the process is $Y$. The projectile is assumed to be moving to the left with total rapidity $Y-Y_0$ (and thus has sizable color charge density $\rho^-$), while the target is moving to the right
 with total rapidity $Y_0$ (and has large $\rho^+$). We assume that the projectile and the target 
contain only partons with large $k^-$ and $k^+$ momenta respectively: $k^->\Lambda^-$ and $k^+>\Lambda^+$. 
The eikonal expression for the $S$-matrix reads
\begin{equation}
{\cal S}_{Y}\,\,=\,\,\int\, D\rho^{+a}(x,x^-)\,\, W^T_{Y_0}[\rho^+(x^-,x)]\,\,\Sigma^P_{Y-Y_0}[\alpha]\,,
\label{s}
\end{equation}
where $\Sigma^P$ is the $S$-matrix averaged over the projectile wave function
\begin{equation}
\Sigma^P[\alpha]\,\,=\,\,\langle P|\,{\cal P} e^{i\int dx^-\int d^2x\hat\rho^{-a}(x)\alpha^a(x,x^-)}\,|P\rangle\,.
\label{sigma}
\end{equation}
($\cal P$ denotes path ordering with respect to $x^-$),
and $W^T[\alpha]$ is the weight function representing the target, 
which is related to the target wave function in the following way: for an arbitrary operator $\hat O[\hat\rho^+]$
\begin{equation}
\langle T|\,\hat O[\hat\rho^+(x)]\,|T\rangle\,\,=\,\,\int\, D\rho^{+a}\,\, W^T[\rho^+(x^-,x)]\,\,O[\rho^+(x,x^-)]\,.
\label{w}
\end{equation}
The field $\alpha(x)$ is the $A^+$ component of the vector potential in the light cone gauge $A^-=0$. 
This is the natural gauge from the point of view of partonic interpretation of the projectile wave function. 
In the same gauge the target charge density $\rho^+$ is related to $\alpha$ through the solution of the classical equations of motion
\begin{eqnarray}
&&\alpha^a(x,x^-)T^a\,\,=\,\,{1\over \partial^2}(x-y)\,
\left\{S^\dagger(y,x^-)\,\,\rho^{+a}(y,x^-)\,T^a\,\,S(y,x^-)\right\}, \nonumber\\
&&S(x,x^-)\,\,=\,\,{\cal P}\,\exp\{i\int_{-\infty}^{x^-} dy^-T^a\alpha^a(x,y^-)\}\,.
\end{eqnarray}
where $T^a_{bc}=if^{abc}$ is the generator of the $SU(N)$ group in the adjoint representation.
The unitary matrix $S(x,x^-\rightarrow\infty)$ as before has the meaning of the scattering matrix of a single gluon from the projectile wave function on the target $|T\rangle$.

These formulae require some explanation. First of all, my notations in this lecture are a little different, since I have reversed the direction of motion of the target and the 
projectile. More importantly, I have reintroduced here the longitudinal coordinate $x^-$. Recall, that in the discussion of the JIMWLK equation I have stressed 
that this coordinate is not necessary. However this was in the situation where I have explicitly assumed the target to be dense. I am not assuming this in the present 
discussion. The variable $x^-$ plays the role of the ordering variable \cite{kl}. The quantities $\rho^{\pm}$ are non commuting operators in their respective Hilbert
spaces. It turns out however that one can trade their non commutativity for an extra coordinate. The value of this coordinate simply traces the position of the 
respective operator in the expression whose expectation value one is calculating. The details do not matter here (those can be found in \cite{kl}), but the upshot is that one can treat
all the charge density operators like classical quantities, if one makes them depend on an extra ``ordering'' variable and calculates their correlation functions
via eq.(\ref{w}).

One can also define an analog of $W^T$ for the wave function of the projectile via
\begin{equation}
\langle P|\,\hat O[\hat\rho^-(x)]\,|P\rangle\,\,=\,\,\int \,D\rho^{-a}\,\, W^P[\rho^-]\,\,O[\rho^-(x,x^-)]\,.
\label{wp}
\end{equation}
With this definition it is straightforward to see that $\Sigma^P$ and $W^P$ are related through a 
functional Fourier transform.
To represent $\Sigma$ as a functional integral with weight $W^P$ we have to order the factors of the charge density $\hat\rho^-$ in the expansion of  
eq.(\ref{sigma}), and then endow the charge density $\hat\rho^-(x)$ with an additional coordinate $t$ to turn it into a classical variable. This task 
is made easy by the fact that the ordering of $\hat\rho$ in eq.(\ref{sigma}) follows automatically the ordering of the coordinate $x^-$ in the path 
ordered exponential. Since the correlators of $\rho(x,t_i)$ with the weight $W^P$ depend only on the ordering of the coordinates $t_i$ and not their 
values, we can simply set $t=x^-$. 
Once we have turned the quantum operators $\hat\rho$ into the classical variables $\rho(x^-)$, the path ordering plays no role anymore, and we thus have
\begin{equation}
\Sigma^P(\alpha)\,\,=\,\,\int\, D\rho^{a}\,\, W^P[\rho]\,\,e^{i\int dx^-\int d^2x\rho^{a}(x,x^-)\alpha^a(x,x^-)}.
\label{sw}
\end{equation}

We now turn to the discussion of the evolution.
The evolution to higher energy can be achieved by  boosting either the projectile or the target. The resulting $S$-matrix should be the same. This is required 
by the Lorentz invariance of the $S$-matrix.
Consider first boosting the projectile by a small rapidity $\delta Y$. 
This transformation leads to the change of the projectile $S$-matrix $\Sigma$ of the form
\begin{equation}
{\partial\over\partial Y}\,\Sigma^P\,\,=\,\,\chi^\dagger[\alpha,{\delta\over\delta\alpha}]\,\,\Sigma^P[\alpha]
\label{dsigma}
\end{equation} 
The explicit form of the kernel $\chi$ is known in the large density limit (JIMWLK) and 
in the small density limit (KLWMIJ \cite{kl} - which we have not discussed here), but is not known in general.
Substituting eq.(\ref{dsigma}) into eq.(\ref{s}) we have
\begin{eqnarray}
&&{\partial\over\partial Y}\,{\cal S}_{Y}=\int \,D\rho^{+a}(x,x^-)\,\, W^T_{Y_0}[\rho^+(x^-,x)]\,\,
\left\{\chi^\dagger[\alpha,{\delta\over\delta\alpha}]\,\,\Sigma^P_{Y-Y_0}[\alpha]\right\}\nonumber\\
&&=
\int \,D\rho^{+a}(x,x^-)\,\,\left \{\chi[\alpha,{\delta\over\delta\alpha}]\,\,
W^T_{Y_0}[\rho^+(x^-,x)]\right\} \,\,\Sigma^P_{Y-Y_0}[\alpha]\,.
\label{ds}
\end{eqnarray}
Where the second equality follows by integration by parts. 
We now impose the requirement that the $S$-matrix does not depend on $Y_0$
\cite{LL}. Since $\Sigma$ 
in eq.(\ref{s}) depends on the difference of rapidities, 
eq.(\ref{dsigma}) implies
\begin{equation}
{\partial\over\partial Y_0}\,\Sigma\,\,=\,\,-\,
\chi^\dagger[\alpha,{\delta\over\delta\alpha}]\,\,\Sigma[\alpha]\,.
\label{dsigma0}
\end{equation} 
Requiring that $\partial{\cal S}/\partial Y_0\,=\,0$ 
we find that $W$ should satisfy 
\begin{equation}
{\partial\over\partial Y}\,W^T\,\,=\,\,
\chi[\alpha,{\delta\over\delta\alpha}]\,\,W^T[\rho^+]
\label{dwt}
\end{equation}

Thus we have determined the evolution of the target eq.(\ref{dwt}) by boosting the projectile
and requiring Lorentz invariance of the $S$-matrix, just as we did in the previous lecture.
On the other hand the extra energy due to boost can be deposited in the target rather than in the projectile. 
How does $W^T$ change under boost of the target wave function? To answer this question
we consider the relation between $\Sigma$ and $W$  together with the evolution of $\Sigma$.
Referring to eqs.(\ref{sw}) and (\ref{dsigma}) it is obvious that multiplication of $\Sigma^P$ by $\alpha$ is 
equivalent to acting on $W^P$ by the operator $-i\delta/\delta\rho$, and acting on $\Sigma^P$ by 
$\delta/\delta\alpha$ is equivalent to multiplying $W^P$ by $i\rho$. Additionally, the action of 
$i\rho$ and $-i\delta/\delta\rho$ on $W^P$ must be in the reverse order to the action of 
$\delta/\delta\alpha$ and  $\alpha$ on $\Sigma^P$. This means that the evolution of the functional 
$W^P$ is given by
\begin{equation}
{\partial\over\partial Y}\,W^P\,\,=\,\,\chi[-i{\delta\over\delta\rho},i\rho]\,\,W^P[\rho]\,.
\label{dwp}
\end{equation}
Although eq.(\ref{dwp}) refers to the weight functional representing the projectile wave function, 
the target-projectile democracy requires that the functional form of the evolution must be the same for $W^T$. 
Comparing eq.(\ref{dwt}) and eq.(\ref{dwp}) we find that the high energy evolution kernel must, as advertised,  satisfy the selfduality relation
\begin{equation}\label{duality}
\chi[\alpha,\,{\delta\over\delta\alpha}]\,\,=\,\,\chi[-i{\delta\over\delta\rho},\,i\rho]\,.
\end{equation}

To stress it again, this mathematical property of selfduality is the reflection of the requirement that the wave functions of the projectile and the target
transform in the same way in their respective frames (and gauges) under the Lorentz transformation.

It is interesting that although we do not know the complete kernel $\chi$, we can check that the known limits of it are consistent with the duality eq.(\ref{duality}).
In particular, the duality transformation interchanges the large and small density limits, and the kernel $\chi$ is known in both these limits. 
In the large density limit the pertinent expression is the JIMWLK kernel. It can be conveniently written as
\begin{eqnarray}\label{JIMWLK1}
&&\chi_{JIMWLK}=
\frac{\alpha_s}{2\pi^2}\int_{x,y,z} {(z-x)_i(z-y)_i
\over (z-x)^2(z-y)^2}\times\\
&&\left\{ {\delta\over \delta \alpha^a(x,0)}{\delta\over \delta \alpha^a(y,0)}+ {\delta\over \delta \alpha^a(x,1)}
{\delta\over \delta \alpha^a(y,1)}- 2{\delta\over \delta \alpha^a(x,0)} S^{ab}(z){\delta\over \delta \alpha^b(y,1)}\right\}\nonumber
\end{eqnarray}
with
\begin{equation}
S^{ab}(z)=
\left[{\cal P}e^{i\int_{0}^{1} d x^-T^c\alpha^c(x^-,z)}\right]^{ab}
\end{equation}
The low density limit of the kernel has recently also been calculated recently \cite{kl}. It reads
\begin{eqnarray}\label{KL}
&&\chi_{KLWMIJ}=-
\frac{\alpha_s}{2\pi^2}\int_{x,y,z} {(z-x)_i(z-y)_i
\over (z-x)^2(z-y)^2}\times\\
&&\{\rho^a(x,0)\rho^a(y,0)+  \rho^a(x,1)\rho^a(y,1)
-2\,  \rho^a(x,0)R^{ab}(z)\rho^b(y,1)\}\,.\nonumber
\end{eqnarray}
with
\begin{equation}
R^{ab}(z)=\left[{\cal P}e^{\int_{0}^{1} d x^-T^c{\delta\over\delta\rho^c(x^-,z)}}\right]^{ab}
\end{equation}
Obviously the kernel eq.(\ref{JIMWLK}) transforms into the kernel eq.(\ref{KL}) by the duality transformation eq.(\ref{duality}). The duality is also present in the dipole limit of the evolution as discussed in \cite{dipoles} and \cite{last}.

Although very interesting. this selfduality property is not powerful enough to unambiguously detemine the structure of the kernel. For example the BFKL kernel is self dual, but clearly is not complete. The same applies to the approximation to the kernel eq.(\ref{smallevolution+}) discussed in \cite{smith} - see \cite{bal1} for discussion of this point. We hope however that once the kernel is known the self duality property will help us to solve the evolution equation.

\section{So long and thanks for all the fish}
So this is where we stand at the moment. We have understood some interesting physics, but there surely is more to come. The immediate future will probably bring more efforts to understand the Pomeron loops and the construction of the complete evolution kernel in the eikonal approximation. This looks like an achievable goal. But then there is of course the question of solving these equations, or at least understanding semiquantitatively the changes that the Pomeron loops bring into the game. Amazingly, some research into the solutions has already been done, even though the equation has not been derived yet \cite{ms},\cite{munier},\cite{levin}. It will be very interesting to see whether these result stay relevant in the framework of the full equation.

I believe there are also other things that can be analyzed in the eikonal approach.
In this lectures I have focused on calculations of the total hadronic cross section. Come to think of it, this is probably the least interesting of all the interesting observables, but it is the simplest one - hence a good starting point. But there has been work also on less inclusive quantities, like calculations of diffractive cross section\cite{kovlev}, one\cite{kovtuch} and two gluon emission probabilities\cite{kovjam},\cite{bknw} within eikonal approach. 
I believe this is the area where the use of the wave function techniques should be most advantageous. The wave function contains much more information than is needed for the total cross section. This additional knowledge is necessary to calculate the various gluonic observables, like one and two gluon emission probabilities and other less inclusive observables. So I hope the techniques discussed in these lectures will find more applications in the not-so-distant future and will prove to be worthwhile developing further.

Finally, I would like to thank Krzysztof Golec-Biernat, Michal Praszalowicz and Andrzej Bialas  for the invitation to give these lectures, and all the participants of the school for making it such an enjoyable experience. I would also like to thank my collaborators over the years and especially Urs Wiedemann and Misha Lublinsky who contributed bulk of the ideas presented here while also cracking some good jokes in the process.
I hope there will be more to come.

For now though, so long and thanks for all the fish.

%

\end{document}